\documentclass[jpcm,graphix,twocolumn,showpacs,10pts]{revtex4-1}
\usepackage[utf8x]{inputenc}
\usepackage{graphicx}
\usepackage{epsfig}
\usepackage{amsmath}
\usepackage{amsfonts}
\usepackage{amssymb}
\begin{document}
\title[]{Normal state electronic properties of LaO$_{1-x}$F$_{x}$BiS$_{2}$ superconductors}

\author{J. D. Querales Flores\footnote{Present address:
Instituto Balseiro, Universidad Nacional de Cuyo, Argentina.},$^{1,2}$  C. I. Ventura,$^{1,3}$  R. Citro$^{4}$ and  J.J. Rodr\'iguez-N\'u\~nez$^{5}$ }
\affiliation{$^{1}$Centro At\'omico Bariloche-CNEA and CONICET, Av. Bustillo 9500, R8402AGP Bariloche, Argentina}
\affiliation{$^{2}$Instituto Balseiro, Univ. Nac. de Cuyo and CNEA, 8400 Bariloche, Argentina}
\affiliation{$^{3}$Sede Andina, Univ. Nac. de R\'io Negro, 8400 Bariloche, Argentina}
\affiliation{$^{4}$Dipartimento di Fisica  ``E.R. Caianiello” and CNR-SPIN, Universit\`a degli Studi di Salerno, I-84084 Fisciano, Italy.}
\affiliation{$^{5}$Lab. SUPERCOMP, Departamento de F\'isica – FACYT, Universidad de Carabobo, 2001 Valencia, Venezuela.}

\date{\today}

\begin{abstract}

A good description of the electronic structure of BiS$_{2}$-based superconductors is essential to understand their phase diagram, normal state and superconducting properties. To describe the first reports of normal state electronic structure features from angle resolved photoemission spectroscopy (ARPES)  in LaO$_{1-x}$F$_{x}$BiS$_{2}$, we used a minimal microscopic model to study their low energy properties. It includes the two effective tight-binding bands proposed by Usui et al [Phys.Rev.B 86, 220501(R)(2012)], 
and we added moderate intra- and inter-orbital electron correlations related to Bi-($p_{Y}$, $p_{X}$) and S-($p_{Y}$, $p_{X}$) orbitals. 
We calculated the electron Green's functions using their equations of motion, which we decoupled in second-order of perturbations on the correlations. We determined the normal state  spectral density function and total density of states for LaO$_{1-x}$F$_{x}$BiS$_{2}$, focusing on the description of the k-dependence, effect of doping, and the prediction of the temperature dependence of spectral properties. Including moderate electron correlations, improves the description of the few experimental ARPES and soft X-ray photoemission data available for LaO$_{1-x}$F$_{x}$BiS$_{2}$. 
Our analytical approximation enabled us to calculate the spectral density around the conduction band  minimum at $\vec{k}_{0}=(0.45\pi,0.45\pi)$, and to predict the temperature dependence of the spectral properties at different BZ points, which might be verified by temperature dependent ARPES.

\end{abstract}


\maketitle

\section{Introduction}
\label{introduction}

In 2012, Mizuguchi et al.\cite{mizuguchi2012} discovered novel layered superconductors which have a crystal structure similar to that of the cuprate and 
iron-based superconductors, the so-called BiS$_{2}$-based superconductors. Experimental and theoretical studies have been carried out in order to
establish the basic properties of these new materials and identify the underlying mechanism for superconductivity.\cite{awana-SSC2013,review2014, review2015} 
Substantial enhancement of superconductivity under moderate pressures in BiS$_{2}$-based compounds has been reported.\cite{awana-JPCS2015} 
Up to date, eleven compounds have been discovered in the BiS$_{2}$ family,\cite{review2014, review2015} and the highest  T$_{c}$ is 10.6 K, reported in LaO$_{0.5}$F$_{0.5}$BiS$_{2}$ under 2 GPa of applied pressure.\cite{mizuguchi2012} 
The crystal lattice of BiS$_{2}$-based superconductors consists of consecutive superconducting BiS$_{2}$ planes and blocking layers. The conduction
planes can be viewed as a square array of Bi atoms, each of them with a basis of two S atoms attached.\cite{mizuguchi2012, review2014, review2015} 
The same crystal structure was found in   LaO$_{0.5}$F$_{0.5}$BiSe$_{2}$,\cite{maziopa2014} in this case 
alternating superconducting BiSe$_{2}$ and blocking LaO layers, with  a smaller  $T_{c}$ = 2.6 K. 

For the discovered superconducting materials within the BiS$_{2}$ family, superconductivity emerges from a metallic normal state\cite{li2013,deguchi2013}
 while the parent compounds are band insulators. In fact, 
 LaO$_{1-x}$F$_{x}$BiS$_{2}$ changes from a band insulator at $x=0$ to a metal 
 by means of chemical substitution: upon electron doping.\cite{mizuguchi2012,li2013,deguchi2013}
In contrast to iron-based superconductors,\cite{cruz2008} in the  LnOBiS$_{2}$ (Ln= La, Ce, Pr, Nd) parent compounds neither magnetic nor structural transitions have yet been reported,  
 indicating that magnetism would be of less relevance to superconductivity in the BiS$_{2}$-based compounds as in iron-based superconductors.\cite{xing2012} 
Recently, the coexistence of superconductivity and ferromagnetism has been reported in Ce$_{1-x}$F$_{x}$BiS$_{2}$ compounds with $x>0.4$, while at $x<0.4$ a paramagnetic  behavior was observed.\cite{demura2015} Very recently, an analysis comparing growth and characterization of LnOBiS$_{2}$ single crystals\cite{nagao2015} highlights the fact 
that sizeable differences may exist between the nominal F-composition (x) and the analytical F-composition (y),  
which is obtained characterizing the samples by X-ray diffraction, scanning electron microscopy and electron probe microanalysis. For instance,   
for LaO$_{1-x}$F$_{x}$BiS$_{2}$ single crystals, to nominal: x = 0.3, 0.4, 0.5, 0.6, respectively correspond analytical values: y= 0.23, 0.37, 0.43. 0.45.\cite{nagao2015}

A number of density functional theory (DFT) band-structure calculations have been reported,\cite{wan2013,usui2012,zhou2013,liu2013, shein2013,suzuki2013,gao2014,yang2013,morice2015,wu2014}
in particular for  LaO$_{0.5}$F$_{0.5}$BiS$_{2}$ and its parent compound.
The valence bands extend from -6 to 0 eV and consist of p-electron states from the O and S atoms, while the bands related to La-$f$, La-$d$ lie far
above the Fermi level ($E_F$), near 4 eV, and the conduction bands are dominated by Bi-$p$ and S-$p$ electron states.\cite{li2013}     
       Based on DFT calculations,  a minimal two-orbital model for LaO$_{1-x}$F$_{x}$BiS$_{2}$ was proposed by  H. Usui et al. \cite{usui2012}
predicting a change of topology of the  Fermi surface (FS)  when electron doping $x$ is increased. Recently, the FS topology proposed for higher doping\cite{usui2012}
 was confirmed by ARPES in LaO$_{0.54}$F$_{0.46}$BiS$_{2}$.\cite{terashima2014}
Usui et al.\cite{usui2012} also suggested the presence of nesting of the FS with wave vector $\vec{k}=(\pi,\pi,0)$ at $x \sim 0.5$, 
due to the quasi-one-dimensional nature of the conduction bands.

Core-level and valence-band soft X-ray photoemission spectroscopy (SXPES)\cite{nagira2014}  were used to investigate the electronic structure  of LaO$_{1-x}$F$_{x}$BiS$_{2}$ ($x=0,0.3,0.5$), which was found to be mainly consistent with  predictions of DFT calculations, including the doping dependence.\cite{yildirim2013,wan2013,li2013}  Nevertheless, a deviation was found between the experimental intensity and spectral shape of the states near  $E_F$ in comparison with those calculated.\cite{yildirim2013,wan2013,li2013} As $x$ increases from 0 to 0.5, a rigid-band shift of the whole bandstructure is predicted, based on the valence band shift observed:  towards higher binding energy by about $\sim0.3$ eV, 
a value which is much smaller than expected from DFT calculations.\cite{yildirim2013} 
Recently, the electronic structure of nearly optimally doped superconductor LaO$_{0.54}$F$_{0.46}$BiS$_{2}$ was investigated using ARPES,\cite{terashima2014} finding 
relatively weak electron correlation effects and a marked influence of spin-orbit coupling on the BiS$_{2}$ planes. A square-like ARPES intensity distribution centered at the BZ center was found, consistent with the prediction of the effective two-band model by Usui et al.\cite{usui2012}, 
who indicated that the Fermi surface was close to a topological transition.

Other theoretical studies have focused on the study of electron-electron correlations and the possibility of unconventional superconductivity in BiS$_{2}$-based compounds.\cite{zhou2013, martins2013, liang2013, yang2013, yao2015} In this context, superconductivity was investigated by Zhou et al.\cite{zhou2013} 
using a microscopic model for LaOBiS$_{2}$ including the effective two-bands  proposed by  Usui et al. \cite{usui2012} and an on-site intra-orbital Coulomb repulsion $U$. Concretely, the spin excitations in the superconducting state were studied calculating the spin susceptibility in random phase approximation (RPA),
and three potential pairing symmetries ( $d$-, $s$- and $p$-wave) were analized, in a weak correlation
scenario: $U\sim0.8 - 2.5$ eV.  With the further addition of an on-site interorbital Coulomb repulsion as well as Hund coupling to the two-orbital model,   Martins et al.\cite{martins2013}  investigated the superconductive pairing properties of BiS$_{2}$-based superconductors.  Employing a weak-coupling multiorbital RPA analysis,  a clear relationship between quasi nesting in the Fermi surface, spin fluctuations, and superconductivity was found,\cite{martins2013} predicting that pairing symmetry measurements may contain a mixture of both  $A_{1g}$ and $B_{2g}$ symmetries, suggesting $U\sim1.08 - 1.8$ eV. In Ref.\cite{liang2014} the pairing symmetry of BiS$_{2}$ compounds was 
investigated  using the two-band model in Ref.\cite{usui2012} by assuming that electron-electron correlation are relevant. It was also assumed\cite{liang2014}  that short range superconductive pairing stems from short antiferromagnetic exchange couplings, finding that extended $s$-wave pairing symmetry is always more favourable than $d$-wave pairing.  
The normal state spectral properties using the two-band minimal model of Ref.\cite{usui2012}, were not investigated in RPA yet.   

Many aspects of the normal state of LaO$_{1-x}$F$_{x}$BiS$_{2}$, such as the effect of electron correlations on the electronic structure, the doping and temperature dependence of the spectral properties,  remain unstudied, which has prompted our investigation. 
Here, we present our study of the paramagnetic normal state of LaO$_{1-x}$F$_{x}$BiS$_{2}$ compounds. In particular, we analize the effect of electron correlations, believed to be moderate in these compounds,  as well as of doping and  of temperature on the spectral properties.  
We do this employing  a minimal microscopic model, which includes the two effective uncorrelated orbitals proposed by H. Usui et al.\cite{usui2012} 
for LaO$_{1-x}$F$_{x}$BiS$_{2}$  as well as intra- and inter-orbital Coulomb interactions,
 and using the method of equations of motion for the Green's functions which we solve in second order of perturbations on the correlations, to determine 
 the temperature, doping and k-dependent spectral density function and density of states for LaO$_{1-x}$F$_{x}$BiS$_{2}$. 
  In a previous work,\cite{condmat2015} we used this analytical approach  to study the normal state spectral properties of ferropnictides,\cite{condmat2015} using for the non-interacting bandstructure the  two effective tight-binding orbitals proposed by Raghu et al.\cite{raghu2008}. Our paper is organized as follows: in Section \ref{section2} we describe the microscopic correlated two-orbital model adopted for our study of the normal state spectral properties of LaO$_{1-x}$F$_{x}$BiS$_{2}$ compounds, and 
the analytical method employed ( which is complemented by Appendix A, where we state the detailed expressions obtained for the relevant Green's functions).    
   In Section \ref{results} we present and discuss our  results at different temperature and doping values, for relevant BZ points, 
  and compare them with the relatively few available theoretical and experimental results for the normal state of LaO$_{1-x}$F$_{x}$BiS$_{2}$. 
In Section \ref{Conclusions} we summarize our main conclusions, prompting for experiments to test our predictions, and mention possible future applications of our work.  
 
\section{Microscopic model and analytical approach.}
\label{section2}

\subsection{Correlated effective model.}
\label{model}

As mentioned in the Introduction, an effective non-interacting two-orbital tight-binding model to describe the low energy physics of LaO$_{1-x}$F$_{x}$BiS$_{2}$, was proposed by Usui et al. in Ref. \cite{usui2012}. Starting from a first-principles band calculation for LaOBiS$_{2}$, where they found that the conduction bands near the Fermi level consist mainly of a mixture of in-plane Bi-6p and S-3p orbitals, they constructed  maximally localized Wannier orbitals to obtain effective tight-binding models, describing the 
kinetic energy part of the effective Hamiltonian. Extracting the portion of the bands which is relevant to the BiS$_2$ layers, assumed to be related to superconductivity,
and neglecting interlayer hoppings, when focusing only on the bands that intersect the Fermi level Usui et al.\cite{usui2012}  obtained a reduced Hamiltonian consisting 
of a two-dimensional effective two-orbital tight-binding model. The model\cite{usui2012} yields two bands (per BiS$_{2}$ layer) which have a quasi-one-dimensional character, providing good nesting of the Fermi surface, with a dispersion characterized by two conduction band  minima\cite{usui2012} located at $X=(\pi,0)$ and $k_{0}=(0.45,0.45)\pi$. 
DFT calculations\cite{usui2012} indicated that the FS topology should undergo considerable changes
with doping as discussed in previous section, which the two-orbital model could describe. 
Usui et al\cite{usui2012} also pointed out that  the total electron filling in the two-orbital model, $n$, should be taken as: $n = x \, $ to describe LaO$_{1−x}$F$_{x}$BiS$_{2}$.  

Concretely, the effective two-orbital tight-binding  model of Usui et al. \cite{usui2012} to describe the kinetic energy part of the effective Hamiltonian for 
LaO$_{1-x}$F$_{x}$BiS$_{2}$,  is given by:

\begin{eqnarray}
  {\mathcal{H}}_{0}  = \sum_{k,\sigma}{\left[{E}_{c}(\vec{k})  \, {c}^{\dagger}_{\vec{k}\sigma} {c}_{\vec{k}\sigma} \,  + \, {E}_{d}(\vec{k}) \, {d}^{\dagger}_{\vec{k}\sigma}{d}_{\vec{k}\sigma}\right]}
  \label{Hamiltonian0}
\end{eqnarray}

where the operators ${c}^{\dagger}_{\vec{k},\sigma}$ and $d^{\dagger}_{\vec{k},\sigma}$ create  an electron in the respective orbitals
$c$ and $d$, with spin $\sigma= \uparrow, \downarrow$, and crystal momentum $\vec{k}$, while the respective energies are:

 \begin{equation}
  E_{  \overset { d }{ c } } (\vec{k}) = \frac{1}{2}\left[\epsilon_{X}+\epsilon_{Y}   \pm \sqrt{ \left(\epsilon_{X}-\epsilon_{Y}\right)^{2} + \epsilon_{XY}^{2}}\right] - \mu
   \end{equation}\label{effectivebands}

where $\mu$ is the chemical potential, and:

\begin{eqnarray}\label{bandas}
 \epsilon_{X}(\textbf{k}) &  = & t_{0} + 2t_{1}(\cos k_{x}+\cos k_{y})\nonumber \\
        &    & + 2t_{3}\cos(k_{x}- k_{y}) + 2t_{4} \cos(k_{x}+ k_{y})  \nonumber \\
      &    & + 2t_{6}\left[ \cos(2k_{x}+k_{y}) + \cos(k_{x}+2k_{y})\right] \nonumber \\
        &    & + 2t_{8}\left[ \cos(2k_{x}-k_{y}) + \cos(k_{x}-2k_{y})\right]  \nonumber  \\
 \epsilon_{Y}(\textbf{k}) &  = & t_{0} + 2t_{1}(\cos k_{x}+\cos k_{y}) \nonumber \\
        &    & + 2t_{3}\cos(k_{x}+ k_{y}) + 2t_{4} \cos(k_{x}- k_{y})  \nonumber \\
      &    & + 2t_{6}\left[ \cos(2k_{x}-k_{y}) + \cos(k_{x}-2k_{y})\right] \nonumber \\
        &    & + 2t_{8}\left[ \cos(2k_{x}+k_{y}) + \cos(k_{x}+2k_{y})\right]  \nonumber  \\
 \epsilon_{XY}(\textbf{k}) &  = & 2t_{2}(\cos k_{x}-\cos k_{y})
 + 2t_{5} (\cos 2k_{x}-\cos 2k_{y}) \nonumber \\
 &  & + 4t_{7}[  \cos(2k_{x}+k_{y}) -  \cos(k_{x}+2k_{y}) ]\nonumber \\
\end{eqnarray}
The hopping parameters up to fourth neighbors in Eqs. \ref{bandas}, between sites on the  square lattice formed by the Bi atoms, 
are taken from Refs. \cite{usui2012,martins2013}:  $t_{0}=2.811$, $t_{1}=-0.167$, $t_{2}=0.107$, $t_{3}=0.880$, $t_{4}= 0.094$, $t_{5}=-0.028$, $t_{6}=0.014$, $t_{7}=0.020$, $t_{8}=0.069$, where all energy parameters are given in eV (as in the rest of our paper).

In our present work, to describe analytically the normal state properties of LaO$_{1-x}$F$_{x}$BiS$_{2}$ compounds, we will consider a minimal model
preserving the essential physics of the problem given by: 

\begin{eqnarray}
 \mathcal{H} & = & \mathcal{H}_{0} + V_{int}
 \label{Hamiltonian}
 \end{eqnarray}
where the kinetic energy is described by $\mathcal{H}_{0}$ of Eq.\ref{Hamiltonian0} including the two effective orbitals proposed by Usui et al.\cite{usui2012} detailed above, 
while we add electron correlations in  $V_{int}$. In fact,  assuming that short-range Hubbard-like electron-electron interactions are present, 
at each site we will consider an intra-orbital Coulomb repulsion $U$, and an inter-orbital repulsion $V$, so that $V_{int}$ reads:

\begin{eqnarray}
V_{int} =  \sum_{i}  \left[ U \left( {n}_{i\uparrow}{n}_{i\downarrow} + {N}_{i\uparrow}{N}_{i\downarrow} \right) 
 + V \left({n}_{i\uparrow}+{n}_{i\downarrow}\right) \left({N}_{i\uparrow}+{N}_{i\downarrow}\right) \right] 
 \label{Vint}
\end{eqnarray}
where: $n_{i\sigma} = {c}^{\dagger}_{i\sigma}{c}_{i\sigma}$ and $N_{i\sigma} = {d}^{\dagger}_{i\sigma}{d}_{i\sigma}$, for spins $ \sigma= \uparrow,\downarrow $.

Thus, we will be modelling LaO$_{1-x}$F$_{x}$BiS$_{2}$ by an effective extended Hubbard model 
consisting of two correlated electron orbitals per site, to study the  electronic properties of the paramagnetic normal state of these compounds: 
in particular,   the k-dependence, effect of doping and temperature dependence of the spectral properties.

\subsection{Analytical calculation of the spectral properties.}
\label{calculations}

To determine the spectral density of $c$ and $d$ electrons in  our effective extended Hubbard model, we calculated the corresponding retarded 
Green's functions at finite temperature introduced by Zubarev\cite{zubarev} defined by:
\begin{align}\label{g-retarded}
G_{\sigma}^{ret}(\vec{k},\omega)  = G(k,\omega + i\delta)  = \ll c_{k \sigma};c^{\dagger}_{k \sigma}\gg (\omega + i\delta) 
\end{align} 
\begin{align}\label{f-retarded}
F_{\sigma}^{ret}(\vec{k},\omega)  = F(k,\omega + i\delta)  = \ll d_{k \sigma};d^{\dagger}_{k \sigma}\gg(\omega + i\delta)
\raisetag{1pt}
\end{align} 
\noindent where $\delta$ is an infinitesimal positive number, and  $\ll \cdots \gg$ represents the usual notation for Zubarev's Green's functions \cite{zubarev} which 
for fermionic operators  $\widehat{A}, \widehat{B}$  are defined as the time Fourier-transform  of the retarded Green's function:
\begin{equation}
 \ll \widehat{A(t)};\widehat{B(t')}    \gg  =  -i \theta(t-t') \, \,  \langle  \, \,  \widehat{A}(t)\widehat{B}(t') + \widehat{B}(t')\widehat{A}(t) \, \,  \rangle 
\end{equation}
 \noindent  where the time-dependent operators appear in Heisenberg representation,  and $ \theta(t) $ is the Heavyside step function.
The  expectation values  $\langle  \cdots \rangle$ of  quantum observables  at  finite temperature are calculated  as the trace of the product of the  density operator 
 and  the observable,  with the  density operator written in terms of the  appropriate  statistical ensemble  at temperature T.\cite{zubarev} 
   In our case, we study the paramagnetic normal state of the system at temperature $T$ with fixed number of electrons $n$,  thus evaluating
 the required expectation values using the grand canonical ensemble.       
  ( The technique can also be extended to study zero-temperature properties, in which case the Green's function definition  involves  
  calculating the expectation values  in the ground state of the system.)

We calculated   $G_{\sigma}(k,\omega)$ and $F_{\sigma}(k,\omega)$   using Zubarev's equations of motion (EOM) formalism,\cite{zubarev,nolting} in particular 
from the following  exact coupled set of equations:  
 
\begin{align}
 & \left[\omega - E_{c}(k) \right] G_{\sigma}(k,\omega)  =  \frac{1}{2\pi} + \sum_{k_{1},k_{2}}\left[\frac{U}{N} {\Gamma}_{1}(k_{1},k_{2},k,\omega)   \right.  \nonumber \\ 
 &\left.   + \frac{V}{N} {\Gamma}_{2} (k_{1},k_{2},k,\omega) + \frac{V}{N} {\Gamma}_{3} (k_{1},k_{2},k,\omega) \right]    \label{eqG}\\
 &\left[\omega - E_{d}(k) \right] F_{\sigma}(k,\omega)  =  \frac{1}{2\pi} +\sum_{k_{1},k_{2}}\left[\frac{U}{N} {\Gamma}_{4}(k_{1},k_{2},k,\omega)  \right. \nonumber \\ 
  & \left. + \frac{V}{N} {\Gamma}_{5} (k_{1},k_{2},k,\omega) + \frac{V}{N} {\Gamma}_{6} (k_{1},k_{2},k,\omega) \right]   \label{eqF} 
\raisetag{1pt}
\end{align}

 \noindent where in Eq. \ref{eqG} we denoted the higher-order Green's functions which encode the charge, spin and orbital fluctuations:   

\begin{eqnarray}\label{set1}
 {\Gamma}_{1}(k_{1},k_{2},k,\omega) & \equiv & {\Gamma}^{ccc}_{\sigma, \overline{\sigma},\overline{\sigma}}(k_{1},k_{2},k,\omega)  \nonumber \\ 
{\Gamma}_{2}(k_{1},k_{2},k,\omega) & \equiv & {\Gamma}^{cdd}_{\sigma, \sigma ,\sigma}(k_{1},k_{2},k,\omega) \nonumber \\
{\Gamma}_{3}(k_{1},k_{2},k,\omega) & \equiv & {\Gamma}^{cdd}_{\sigma, \overline{\sigma} ,\overline{\sigma}}(k_{1},k_{2},k,\omega)
\end{eqnarray}

 \noindent and in Eq. \ref{eqF}: 
\begin{eqnarray}\label{set2}
 {\Gamma}_{4}(k_{1},k_{2},k,\omega) & \equiv & {\Gamma}^{ddd}_{\sigma, \overline{\sigma},\overline{\sigma}}(k_{1},k_{2},k,\omega) \nonumber \\
{\Gamma}_{5}(k_{1},k_{2},k,\omega) & \equiv & {\Gamma}^{dcc}_{\sigma, \sigma ,\sigma}(k_{1},k_{2},k,\omega) \nonumber \\
{\Gamma}_{6}(k_{1},k_{2},k,\omega) & \equiv & {\Gamma}^{dcc}_{\sigma, \overline{\sigma} ,\overline{\sigma}}(k_{1},k_{2},k,\omega)
\end{eqnarray}

\noindent while the following notation was introduced above: 
\begin{equation}\label{definition}
  \Gamma^{\alpha,\beta,\gamma}_{\sigma_{\alpha}\sigma_{\beta}\sigma_{\gamma}}(k_{1},k_{2},k,\omega) \equiv 
    \, \,  \ll c_{\alpha_{k_{2}, \sigma_{\alpha}}} c^{\dagger}_{\beta_{ k_{1}, \sigma_{\beta}}} c_{\gamma_{k_{1}-k_{2}+k,\sigma_{\gamma}}}  ;  c^{\dagger}_{k \sigma}\gg (\omega)\\ 
\end{equation}

The Hartree-Fock solution to this problem is obtained by closing the set of equations of motion in first-order of perturbations on the correlations, 
performing a mean-field approximation  of the $\Gamma_{i}$ $(i = 1,6)$ Green's functions in Eqs.\ref{eqG} and \ref{eqF},  by which each is expressed   
only in terms of  $G_{\sigma}(k,\omega)$ and $F_{\sigma}(k,\omega)$. In Appendix A.1 we state the Hartree-Fock solutions for the  $c$ and $d$ electron Green's functions.

In the present  work, instead, we solved the problem at a higher level of approximation: 
we went on to calculate the three exact equations of motion for the $\Gamma_{i}$ $(i = 1,3)$ Green's functions,  coupled to  $G(\vec{k},\omega)$ in first order 
through Eq.\ref{eqG}; and proceeded likewise for $\Gamma_{i}$ $(i = 4,6)$ coupled to $F(\vec{k},\omega)$ through Eq.\ref{eqF}. 
Each of the six second order equations of motion for $\Gamma_{i}$ $(i = 1,6)$ introduces three new coupled  higher-order Green functions in the problem.

To close the coupled set of equations of motion in second-order of perturbations on the correlations, 
 we used the approximation  previously employed in our investigation of ferropnictides in Ref.\cite{condmat2015} (for full details 
 of our calculations, see Appendix A of Ref.\cite{condmat2015}). Briefly, our approach consists in approximating 
all the higher-order Green's functions introduced in each subset of equations of motion for $\Gamma_{i}$ $(i = 1,3)$, 
  expressing them in mean-field only in terms of $\Gamma_{i}$ $(i = 1,3)$  and $G(\vec{k},\omega)$.
We proceeded likewise for the subset of equations for  $\Gamma_{i}$ $(i = 4,6)$ related to  $F(\vec{k},\omega)$. 
This second-order  approximation yields a closed system of eqs. of  motion for $G$,  $\Gamma_{1}$, $\Gamma_{2}$, $\Gamma_{3}$; 
and an analogous one for $F$,  $\Gamma_{4}$, $\Gamma_{5}$, $\Gamma_{6}$, which we solved to determine $G(\vec{k},\omega)$ and $F(\vec{k},\omega)$ 
obtaining the expressions presented in Appendix A.2. Notice that the evaluation of $c$ and $d$ Green's functions obtained in our second-order 
perturbative approach requires performing double and triple summations over the first Brillouin zone (BZ) of the crystal lattice.
For simplicity, we have assumed a square lattice, as done previously,\cite{usui2012}  and for the BZ summations 
used the Chadi-Cohen BZ sampling method.\cite{macot} As well known for other Hubbard-like models (see e.g. Refs. \cite{roth1969,nolting}), the decoupling scheme 
adopted for the higher order Green's functions coupled in the second-order EOM, accounts for non-trivial k-dependent self-energy effects 
of the single-particle Green function, as our results in next section will show (and we previously demonstrated for ferropnictides: 
in particular, in Appendix B of Ref.\cite{condmat2015}).


With the $c$ and $d$ electron Green's functions determined for the paramagnetic normal state of our effective extended Hubbard model,
we calculate  the total spectral density function, to be compared with ARPES experiments, $A(\vec{k},\omega)$, 
as follows:   

\begin{equation}\label{partial}
A(\vec{k},\omega) = \sum_{\sigma} \,  A_{c\sigma}(\vec{k},\omega) + A_{d\sigma}(\vec{k},\omega)
\end{equation}
where: 

\begin{equation}
 A_{c\sigma}(\vec{k},\omega)  =  -\frac{1}{\pi} Im G^{ret}_{\sigma}(k,\omega) ;  \, \, \, \, \, A_{d\sigma}(\vec{k},\omega)  =  -\frac{1}{\pi} Im F^{ret}_{\sigma}(k,\omega)
 \label{partialdos}
 \end{equation}

The total density of states (DOS), denoted as $A(\omega)$, is obtained by integrating over the whole BZ the spectral density function: 
\begin{equation} 
A(\omega) = \sum_{k \in BZ} \,  A(k,\omega)  \quad 
\label{totalA}  
\end{equation}

Finally, the self-consistent set of coupled equations to be solved in our second-order perturbative approach is completed by the equation which yields  
the total electron band filling $n$ (or chemical potential $\mu$) at temperature $T$:  

\begin{align}\label{fill}
  n (\mu) = x = n_{c} + n_{d} = \int \frac{d\vec{k}}{(2\pi)^{2}} \frac{d\omega}{2\pi} \left[ A_{c}(\vec{k},\omega) \frac{1}{e^{\beta  E_{c}(\vec{k})} +1 }  \right. \nonumber \\
    \left. + A_{d}(\vec{k},\omega) \frac{1}{e^{\beta E_{d}(\vec{k})} +1 } \right]
\raisetag{1.5pt}
\end{align}
where $\beta = \frac{1}{k_{B}T}$ and $k_{B}$ is Boltzmann's constant. 


\section{Results and discussion}
\label{results}

In this section we present the electronic structure results obtained in our approach for the paramagnetic normal
state of LaO$_{1-x}$F$_{x}$BiS$_{2}$ superconductors, at different temperature and doping values, using the model presented in Section \ref{model}. 
We compare our results  with the relatively few experimental ARPES and soft X-ray photoemission data and previous theoretical results which are  available for these compounds. 
 We solved the self-consistent equations for the Green's functions  
detailed in Section \ref{calculations} numerically, using the tight-binding parameters of Ref. \cite{usui2012} for the non-interacting effective bands 
proposed to describe LaO$_{1-x}$F$_{x}$BiS$_{2}$.  
Regarding the precision of the Chadi-Cohen sampling method \cite{macot} used for the evaluation of Brillouin zone summations in the Green's functions, 
it is specified by the order $\nu$:  
we found good convergence in most results using the seventh order ($\nu=7$) for the square-lattice BZ Chadi-Cohen sampling,\cite{macot} but in this paper 
 all spectral density function and total density of states results presented 
 were obtained  using ninth order for improved accuracy, except for Figs. 1 and 2 where $\nu=8$ was used (as indicated in the respective figure captions). Notice  
 that $\nu= 8$ order for the square-lattice BZ Chadi-Cohen sampling implies using 8,256 special BZ points, while $\nu = 9$ implies using 32,896 ones.\cite{macot} For the retarded Green's functions  $\delta= 10^{-6}$ was used in the present work, except for the DOS results plotted: with $\delta=10^{-4}$ (these are typical values, for instance in Ref.\cite{martins2013}  $\delta=10^{-5}$ was used).

We present our results divided in subsections, each of them focusing on different aspects of the problem. First, we discuss the 
separate effect of the intra- and inter-orbital electron correlations in our approach for LaO$_{1-x}$F$_{x}$BiS$_{2}$, in particular on the density of states. We also  analize 
the renormalization by correlations of the Fermi surface and effective bands. 
Comparison with chemical potential shift data for specific electron doping, allow us to identify the range of moderate correlation values 
to be used in the rest of our paper for a proper description of the compounds.    
Next, in subsection \ref{k-effect} we focus on the momentum 
dependence of the spectral density function along symmetry paths of the square lattice Brillouin zone, 
comparing our results with the available reported ARPES data at X, and predicting the spectral density to be expected near $k_0= (0.45,0.45)\pi$ (the second relevant conduction band minimum for LaO$_{1-x}$F$_{x}$BiS$_{2}$).  In subsection \ref{doping-effect} we discuss the doping dependence of the total density of states. In subsection \ref{T-effect} we predict the temperature dependence of the total density of states and of the spectral density function at different k-points for LaO$_{1-x}$F$_{x}$BiS$_{2}$, and compare with 
available data on NdO$_{1-x}$F$_{x}$BiS$_{2}$.

\subsection{Effect of intra- and inter-orbital Coulomb interactions on the electronic structure.}      
 \label{UV-effect}     
      
 Though several scenarios have been proposed to describe the superconductivity in BiS$_{2}$-based compounds, as discussed in Section \ref{introduction}, 
     there is still no agreement  on the relevant mechanism originating it. Similarities with the cuprate and pnictide superconductors indicate the possibility of an unconventional pairing mechanism involving electron correlations,\cite{martins2013} possibly not strong.\cite{usui2012}

In this subsection we analize the effect on the electronic structure of the intra- and inter orbital correlations included in the model for LaO$_{1-x}$F$_{x}$BiS$_{2}$ presented in Section \ref{model}.
We compare the results obtained with our second order perturbative approach for the electron Green's functions, detailed in Section \ref{calculations} and  \ref{apen1}, 
 with other theoretical and the relatively few experimental data yet available, determining typical values for $U$ and $V$ to be used in the rest of the paper.

     \begin{figure}[h!]
  \begin{center}
  \includegraphics[width=8cm]{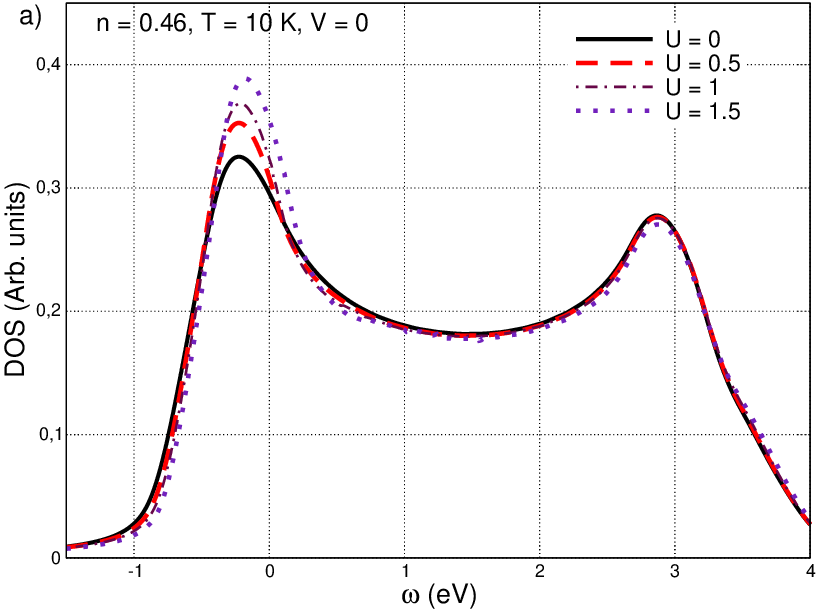}
  \includegraphics[width=8.cm]{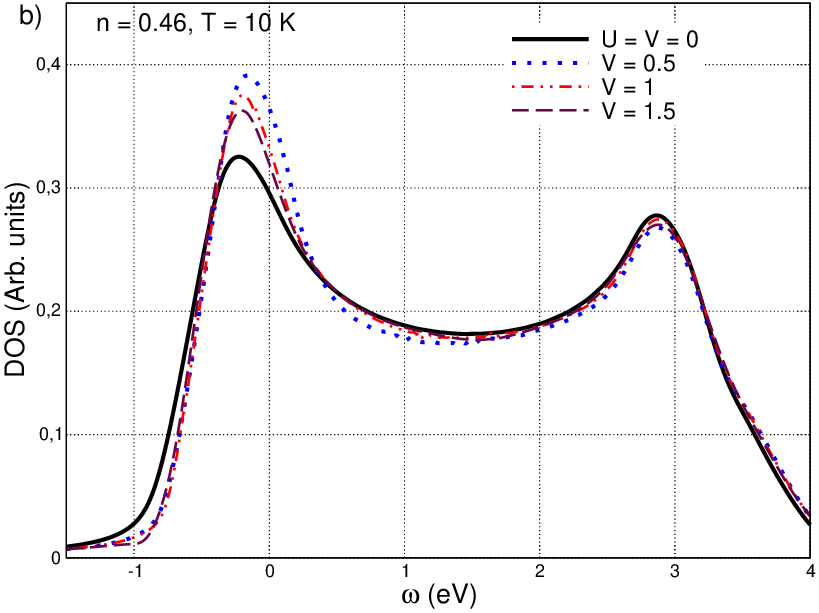}
   \includegraphics[width=7cm]{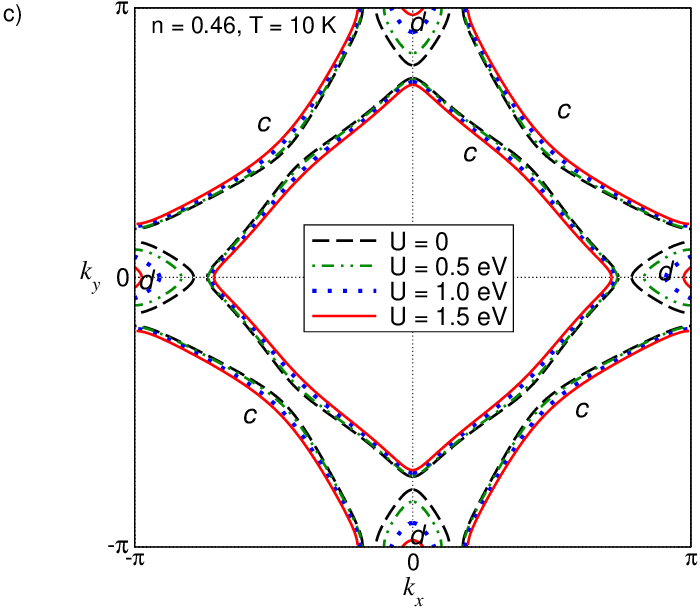}
    \caption[]{a) Total density of states  as a function of energy,  $A (\omega)$ ($\omega$ measured w.r. to the Fermi level, $E_F$),
     for different values of  $U$ (as  indicated in the plot), and at fixed $V=0$;
    b)  $A (\omega)$  for different values of V ( as indicated in the plot), at fixed $ U = 1.5 eV$ (except for the solid line: depicting 
    the uncorrelated case); c) Effect of correlations on the FS: $U$ as indicated in plot and: $ V = U/2$, $c$ and $d$ labels indicate 
    the orbital character of each set of Fermi surface sheets.  
     In this figure:  electron band filling $ n = x = 0.46$ ( $\mu  = 1.11$ eV) and  temperature: $T = 10$ K. Non-interacting tight-binding parameters
      from Ref. \cite{usui2012}:  $t_{0}=2.811$, $t_{1}=-0.167$, $t_{2}=0.107$, $t_{3}=0.880$, $t_{4}= 0.094$, $t_{5}=-0.028$, $t_{6}=0.014$, 
     $t_{7}=0.020$, $t_{8}=0.069$, and in the units of eV.  $\vec{k}_{0}=(0.45,0.45)\pi$ ( in eV).  Chadi-Cohen BZ summations order: $\nu= 8$.}
  \label{dep-U}
  \end{center}
  \end{figure}

We first show the separate effects of the intra- and inter-orbital correlation parameters included in the model.   
The total density of states  as a function of energy is shown in Fig.  \ref{dep-U}(a), for different values of intra-orbital correlation $U$ between 0 and 1.5 eV,
at fixed inter-orbital: $V = 0$, and at the same temperature $T = 10$K and electron filling $n=0.46$ 
as in the only normal state ARPES experiment yet available for the LaO$_{1-x}$F$_{x}$BiS$_{2}$ family\cite{terashima2014}.  
At those temperature and filling values, we have for the chemical potential: $\mu = 1.11$ eV.
  Being the total bandwidth $W\sim4.5$ eV for the non-interacting (bare) 
effective band structure of Ref.\cite{usui2012},  $U=1.5$ eV would correspond approximately to $ 0.33 W $.
Increasing $U$, we predict that the main difference in the DOS curves (see Fig. \ref{dep-U}(a)) would be a monotonous increase of the spectral weight 
of the peak located slightly below $E_F$ w.r. to the non-interacting case ($U=0=V$).   

 \begin{figure}[t!]
  \begin{center}
  \includegraphics[width=8.6cm]{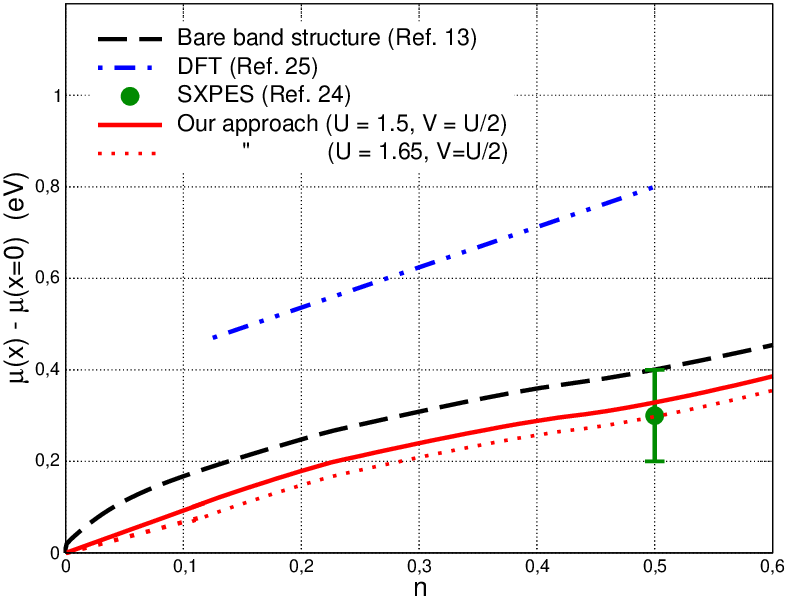}
    \caption[]{Dependence of the chemical potential shift on total electron band filling ( n = x ): comparison between  soft X-ray photoemission data at T= 300 K
    in LaO$_{1-x}$F$_{x}$BiS$_{2}$,\cite{nagira2014}, DFT predictions,\cite{yildirim2013} the non-interacting two-orbital model \cite{usui2012} 
    and the results including correlations with our second-order perturbative approach at  T= 300 K.}
      \label{shift}
  \end{center}
  \end{figure}

In Fig.  \ref{dep-U}(b), we focalize on the effect of the inter-orbital Coulomb  interaction $V$ on the density of states, when a fixed $U = 1.5 eV $ is considered 
(the $U=0=V$ case is included for comparison).  We find that the effect of increasing $V$ is opposite to that of increasing $U$, effectively tending to screen it: 
the main effect of increasing $V$ is a monotonous decrease of the spectral weight of the DOS peak located slightly below the Fermi level, 
 evolving the resulting DOS  towards the non-interacting one ($U=0=V$). Therefore, due to the differences in the relevant non-interacting effective bandstructure 
of LaO$_{1-x}$F$_{x}$BiS$_{2}$\cite{usui2012} and  ferropnictides \cite{raghu2008}, and the corresponding differences in location of the Fermi levels, 
we predict a striking difference in the effect of the inter-orbital electron correlation $V$: which is relevant in BiS$_{2}$-based superconductors, 
while largely irrelevant for ferropnictides\cite{condmat2015,dagotto2011,scalapino2012}. This is a fact already pointed out in connection with other correlated systems,\cite{hague,yamada}, where the presence of two or more bands close in energy to $E_F$,  with relatively low density of electrons ( as  would be the case  for LaO$_{1-x}$F$_{x}$BiS$_{2}$\cite{usui2012}), would amplify the effect of inter-orbital electron correlations.


The non-interacting Fermi surface of LaO$_{1-x}$F$_{x}$BiS$_{2}$ has been discussed in Ref.\cite{usui2012,martins2013}, 
it includes  a topological FS change between  one topology characteristic of low doping, 
experimentally confirmed by ARPES in related compounds: NdO$_{1-x}$F$_{x}$BiS$_{2}$\cite{zeng2014,ye2014},
 and a FS topology characteristic of high  doping, experimentally confirmed for LaO$_{0.54}$F$_{0.46}$BiS$_{2}$ by ARPES\cite{terashima2014}.
 Using the non-interacting effective two-orbital model by Usui et al.\cite{usui2012},  the FS change of topology would take place at the 
 critical doping: $ n_{crit} = 0.44$, while we checked that this value is reduced when correlations are taken into account: 
 for $ U = 1.5 eV = 2 V $ with our second-order perturbative approach we find $ n_{crit} = 0.35 $ , while in Hartree-Fock approximation $ n_{crit} = 0.22$.
In  Figure \ref{dep-U}(c) we exhibit the effect of the renormalization by correlations on the FS  obtained in our approach for LaO$_{0.54}$F$_{0.46}$BiS$_{2}$ 
and compare it with the non-interacting result\cite{usui2012},  at  temperature $T=10K$ for the nominal composition $x = n = 0.46$, exhibiting 
a FS with the characteristic high doping topology. 
We predict that the main effect of increasing electron correlations would be to enhance the nesting properties of large portions of the FS (in particular, 
  of the large central sheet around $\Gamma$, related to $c$ orbitals). Interestingly, a similar effect on nesting was reported by Martins et al.\cite{martins2013} 
  but  as a result from an increase of doping  in the same paper where they established a relationship between the quasi-nesting properties of the FS, 
  spin fluctuations and superconductivity in BiS$_2$ superconductors.  
 In Fig.\ref{dep-U}(c) we observe that the increase  of electron correlations increases the area of the pockets related to the $d$-band, to compensate for the 
 reduction of the FS area related to the $c$ band.

In order to determine typical values for $U$ and $V$  for an adequate description of LaO$_{1-x}$F$_{x}$BiS$_{2}$, 
in Figure \ref{shift} we compare different results for the chemical potential shift as a function of doping.  We include the experimental SXPES estimation 
of Ref.\cite{nagira2014}  who, assuming that the Fermi level of the $x=0$ sample is located at the bottom of the conduction band, deduced a chemical potential shift of $\sim0.3$eV between the $x=0$ and $0.5$ samples. This value is much smaller than the expected one from band calculations in Ref.\cite{yildirim2013} 
(0.8 eV), also plotted. Notice in Figure \ref{shift}, also,  that the non-interacting effective two-orbital model\cite{usui2012} yields a chemical potential shift of $\sim0.4$ eV between $x$=0 and $0.5$, and that adding correlations with our perturbative approach, the description of the experimental results can be improved.  
Therefore, in the following   we will present our results for the description of LaO$_{1-x}$F$_{x}$BiS$_{2}$  fixing  $U=1.5$ eV and $V = 0.75$ eV,
 similar values to those used in Ref.\cite{martins2013} to study other properties. 

     \begin{figure}[h!]
  \begin{center}
  \includegraphics[width=8.6cm]{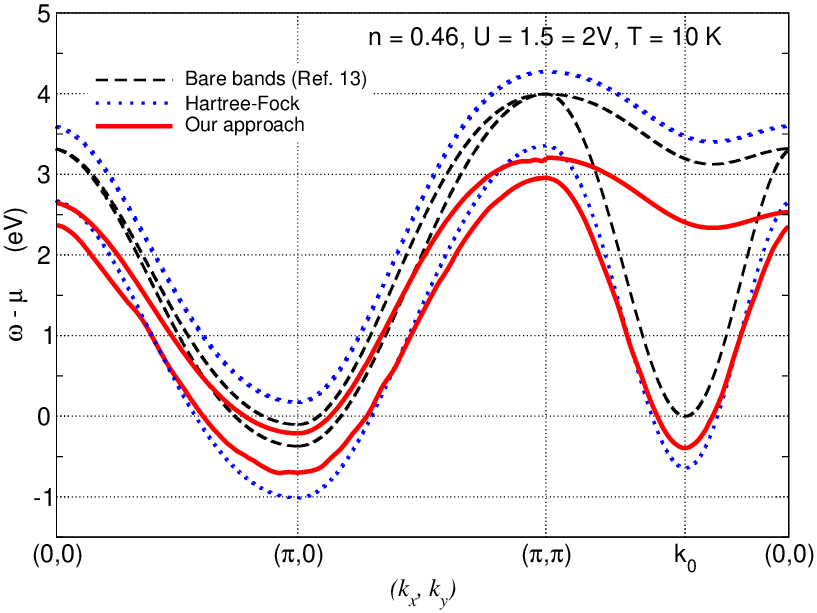}
    \caption[]{Band structure renormalization: comparison of the renormalization by correlations (  $ U = 1.5 eV = 2 V $. Solid lines: our second-order perturbative approach; dotted lines: Hartree-Fock) with  the non-interacting (dashed lines \cite{usui2012}) band structure of the two-orbital model for LaO$_{1-x}$F$_{x}$BiS$_{2}$.   In each case:  the upper band corresponds to the $d$-orbital, and the lower band to the $c$-orbital.  $\vec{k}_{0} = (0.45,0.45)\pi$.
    Other parameters as in Fig.1, except for:  $ \nu = 9 $.}
  \label{renorbands}
  \end{center}
  \end{figure}

Finally, in Figure \ref{renorbands} we compare the renormalized band structure for LaO$_{1-x}$F$_{x}$BiS$_{2}$  obtained including correlations in our second-order perturbative approach (detailed in Section \ref{calculations} and \ref{apen1})  with the first-order Hartree-Fock results, 
and  the effective bands of  the non-interacting two-orbital  model proposed in Ref.\cite{usui2012}. 
Notice that, by including correlations, at $\Gamma$ and $M$  the degeneracy of the non-interacting band structure is broken, 
 as a result of the difference in the  c- and d-band self-energies obtained in both the  first order(Hartree-Fock)  
 and the second order perturbative approaches  whose results are detailed in \ref{apen1}, i.e. 
 in both approximations an orbital-dependent renormalization by correlations appears, which gives rise to the  lifting of the degeneracy at $\Gamma$ and $M$.
While in Hartree-Fock approximation  the rigid-band (k-independent) renormalization yields a larger total bandwidth, 
the renormalization obtained with our second-order approximation is k-dependent and  yields a narrowing of the total band structure w.r. to the non-interacting two-orbital model.\cite{usui2012} Notice  in Figure \ref{renorbands}, also,  that while  the non-interacting model and our second-order perturbative approach to include correlations 
agree near $\vec{k} = (\pi,0)$, in placing  electrons in the  $c$ and $d$ bands which cross the Fermi level, in Hartree-Fock only $c$-electrons appear; 
while at $\vec{k}_{0} = (0.45,0.45)\pi$: all three approaches agree in that only $c$ electrons appear, though the non-interacting band barely touches  $E_F$ while 
more $c$ electrons are predicted to be present when correlations are included.

\subsection{Momentum dependence of the spectral density $A(\vec{k},\omega)$} \label{k-effect} 
    
       In this section, we exhibit the spectral density function results obtained with our approach  along two BZ paths, around the two relevant minima        of the  LaO$_{1-x}$F$_{x}$BiS$_{2}$ bandstructure, shown  in Figure \ref{renorbands}.  
       
       \begin{figure}[h!]
  \begin{center}
  \includegraphics[width=8.0cm]{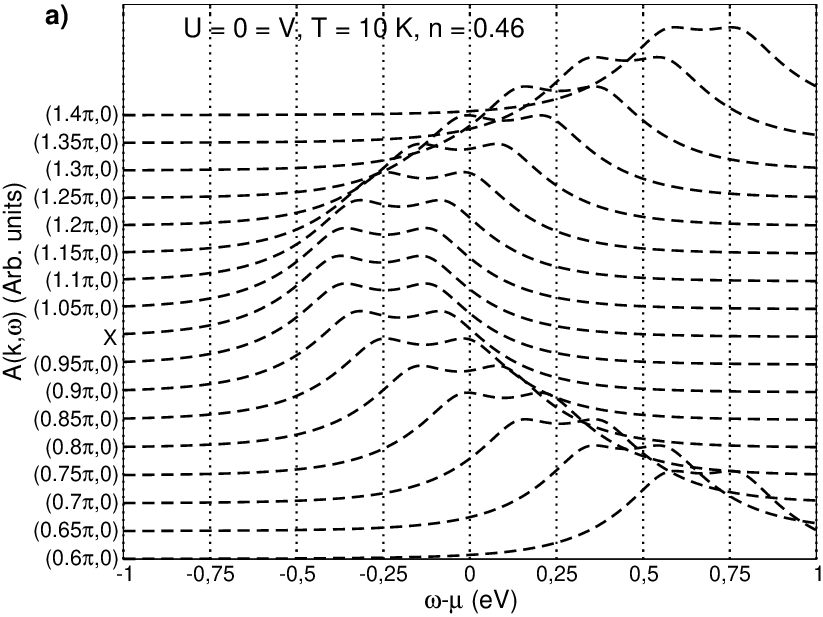}
  \includegraphics[width=8.0cm]{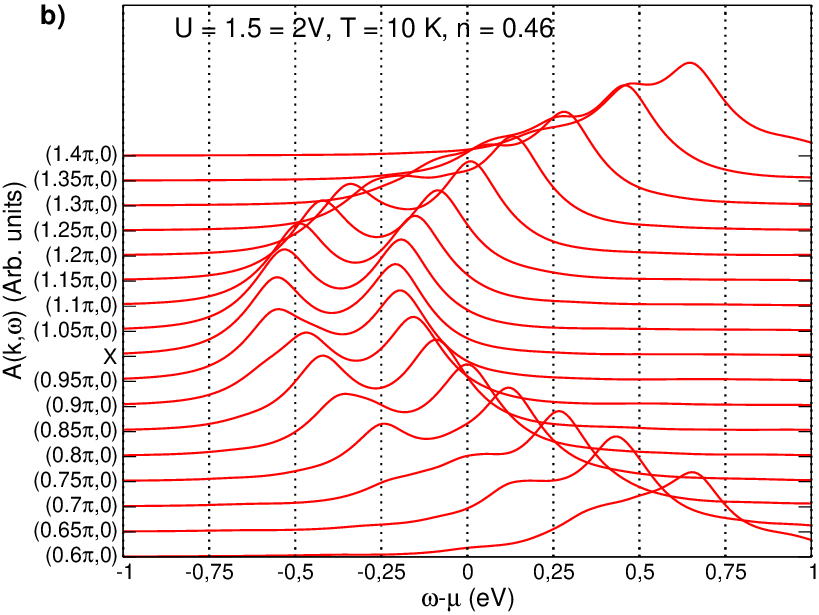}
    \caption[]{ Spectral density $A(\vec{k},\omega)$ as a function of energy $\omega$, for differents BZ points  (shown vertically desplaced along the symmetry path $X-\Gamma$: a) for the non-interacting two-orbital model \cite{usui2012},  b) present work: correlated two-orbital model, with our second-order perturbative approach.
      Parameters: $U = 1.5 eV = 2 V $, Other parameters as in Fig.1, except for:  $ \nu = 9 $.}
  \label{AkwX}
  \end{center}
  \end{figure}

           First, in Fig. \ref{AkwX} we exhibit $A(\vec{k},\omega)$ as a function of crystal momentum $\vec{k}$ around $X$, at low temperature: $T=10$K, and doping $n=x=0.46$, parameters chosen as to compare our results directly with the only available ARPES data for La$_{1-x}$F$_{x}$BiS$_{2}$ compounds.\cite{terashima2014}  Fig. \ref{AkwX}(a) shows
             the spectral density function calculated with the non-interacting two-orbital model of Usui et al.\cite{usui2012}: 
             notice that along the BZ path considered around $X$ it exhibits a two-peak structure. The two well-defined and separated peaks, possessing equal spectral weight,  
           represent the non-interacting(bare)  $c$ and $d$  bands crossing $E_F$ near X shown in Fig. \ref{renorbands}. 
            ARPES experiments for LaO$_{0.54}$F$_{0.46}$BiS$_{2}$\cite{terashima2014} indicate that the bottom of the lower conduction band $E_{min}$ is located near - 0.75 eV, 
            while in Fig. \ref{AkwX}(a) and Fig.\ref{renorbands}, in the absence of correlations it is placed near - 0.37 eV.
            Interestingly, including renormalization effects by correlations with our second-order perturbative approach 
            we find that the agreement with ARPES improves: 
            concretely,  in Fig.  \ref{AkwX}(b) we exhibit our calculated  renormalized spectral density function around $X$, 
            yielding $E_{min} \sim - 0.7 eV$ for $ U = 1.5 eV = 2 V $ (as also noticeable in Fig.3), closer to the ARPES result.
               Meanwhile,  notice that the Hartree-Fock approximation overestimates the correction, yielding $E_{min} \sim  - 1 eV$, and 
               equal spectral weight for the two peaks.  
               As seen in Fig. \ref{AkwX}(b), our approach including correlations also captures the  
               shift in spectral weight distribution between the two peaks along the BZ path, which is present in ARPES results.\cite{terashima2014}


     \begin{figure}[h!]
  \begin{center}
  \includegraphics[width=8.0cm]{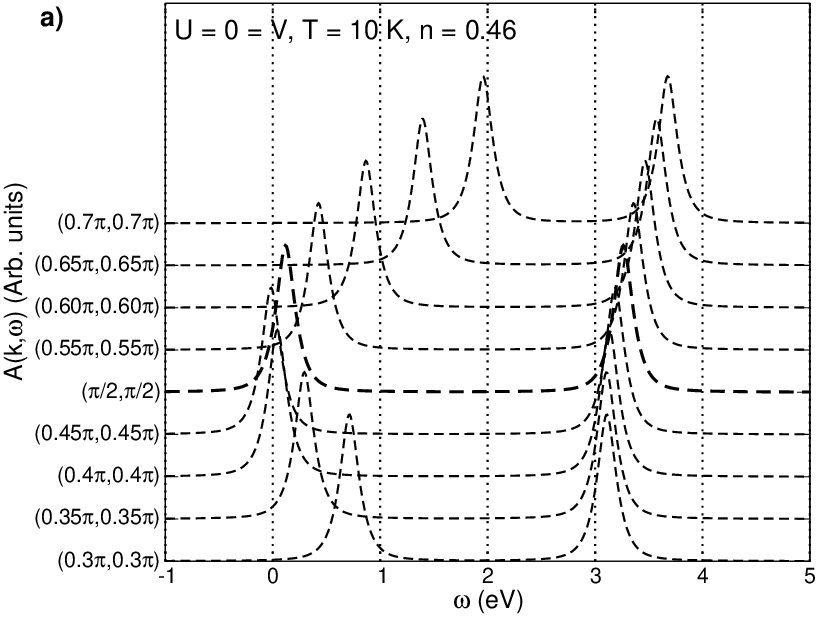}
  \includegraphics[width=8.0cm]{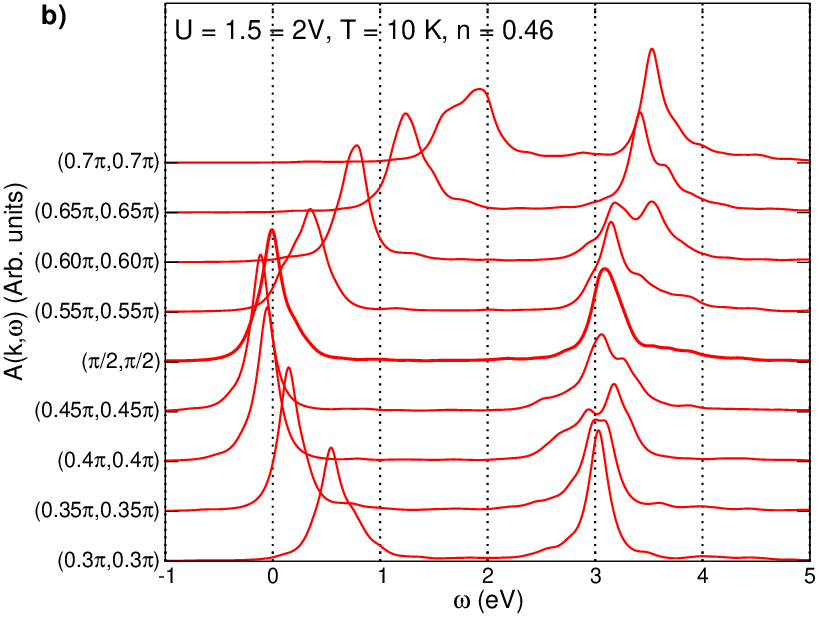}
    \caption[]{ Spectral density $A(\vec{k},\omega)$ as a function of energy $\omega$, for differents BZ points  (shown vertically desplaced along the symmetry path $M-\Gamma$, around $\vec{k}_{0}$: a) for the non-interacting two-orbital model \cite{usui2012},  b) present work: correlated two-orbital model, with our second-order perturbative approach. 
    Parameters: $U = 1.5 eV = 2 V $, other parameters as in Fig.1, except for:  $ \nu = 9 $.}
  \label{Akw-min2}
  \end{center}
  \end{figure}

 As a prediction,  in Figures \ref{Akw-min2}(a) and \ref{Akw-min2}(b) we exhibit the non-interacting spectral density  and our calculated renormalized spectral density function   along the symmetry path $M-\Gamma$, around $\vec{k}_{0} = (0.45\pi,0.45\pi)$, respectively. This BZ region has not yet been probed by ARPES for LaO$_{1-x}$F$_{x}$BiS$_{2}$ compounds.  Comparing the results in Fig.\ref{AkwX}(b) with those in Fig.\ref{Akw-min2}(b), one sees that the renormalization effects by correlations around $E_F$ 
 are more relevant near $X$ than near  $\vec{k}_0$.


\subsection{Effect of electron doping}\label{doping-effect}
    
\begin{figure}[h!]
  \begin{center}
  \includegraphics[width=7.0cm]{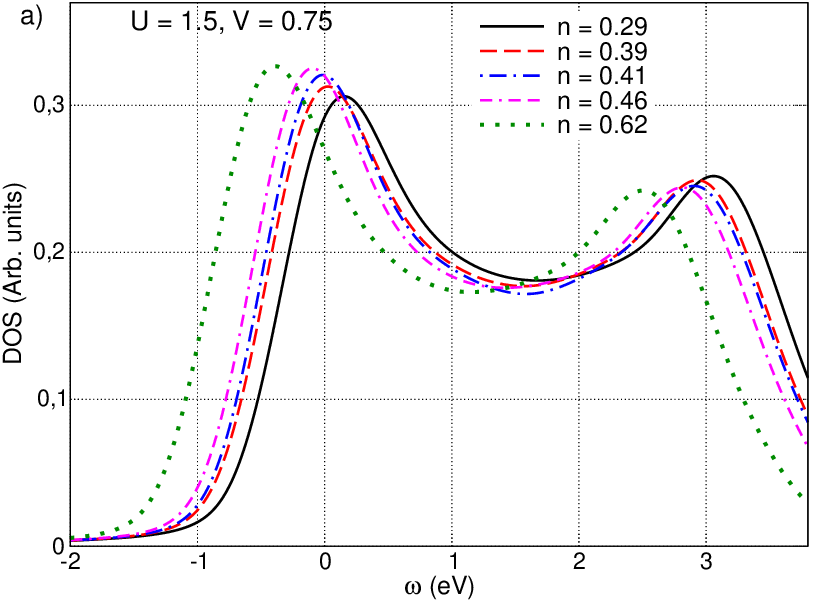}
  \includegraphics[width=6.0cm]{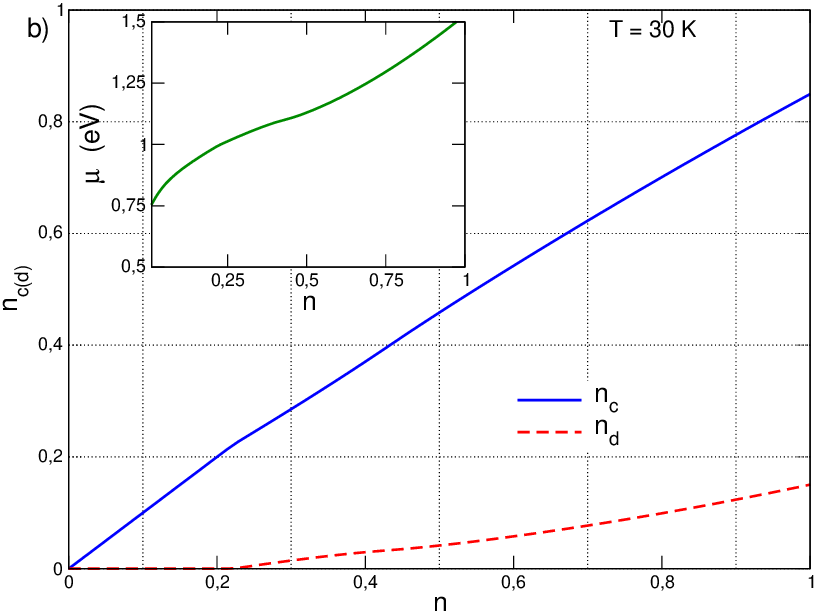}
    \caption[]{ Doping effect. (a) DOS $A(\omega)$, for different values of electron doping (indicated in the figure) at $T = 10$K. (b) 
Filling of the $c$ and $d$ bands, as a function of doping, at $T = 30 K$.   Inset: dependence of the chemical potential on total band filling at $T = 30 K$.   
    $U=1.5 eV = 2 V$, and other parameters as in Fig.1, except for:  $ \nu = 9 $. }
  \label{ef-doping-Aw}
  \end{center}
  \end{figure}

  In this subsection we focus on effects of electron doping  on the spectral properties.
In Figure \ref{ef-doping-Aw}(a), we show our calculated total density of states for doping values $n$ between $0.29$ and $0.62$. Increasing electron-doping, $n$, we find 
an almost rigid-band-like shift of the peak positions to lower energies, in agreement with results of SXPES\cite{nagira2014}  and DFT\cite{yildirim2013} near the Fermi level in LaO$_{1-x}$F$_{x}$BiS$_{2}$ with x between 0 and 0.5. 

Our DOS results also describe an increase upon doping of the spectral weight of the peak located at lower energy (closest to $E_F$), 
  as reported  in the SXPES experiment.\cite{nagira2014} This fact is reminiscent of the mentioned effect of the increase of $U$ on the DOS and on the FS topology 
  discussed in Section \ref{UV-effect}, also in connection with Ref.\cite{martins2013}, and would justify the reduction upon increase of $U$  
  of $n_{crit}$ for the FS topology change. Nevertheless, in Figure \ref{ef-doping-Aw}(a) we find a difference: because 
  the increase of doping not only increases the spectral weight of the peak near $E_F$, as increasing $U$ does, but it also shifts the peaks to lower energies 
  as found in SXPES.\cite{nagira2014}

 As mentioned before, the effective two-orbital model,\cite{usui2012} which was determined taking into account mainly the Bi-$6p$ and S-$3p$ orbitals near $E_F$, is suitable to describe 
   the low-energy properties of La$O_{1-x}$F$_{x}$BiS$_{2}$. For a description of the DOS between 1 eV and 4 eV, the inclusion of La and O orbitals would be important.\cite{yildirim2013, review2015} 
   
    In Figure  \ref{ef-doping-Aw}(b), we exhibit the effect of doping on the filling of each effective band: though both show a monotonous increase of filling with $n$,
     as mentioned before most electrons occupy the lower $c$ band. We find  the $d$ band almost empty below $n = 0.2$.  
     The inset of Figure \ref{ef-doping-Aw}(b) shows the chemical potential as a function of $n$ at low temperature $T = 30 K$, exhibiting an inflection point at $n= 0.44$.

\subsection{Effect of temperature on the DOS and the spectral density}

\label{T-effect}

\begin{figure}[h!]
  \begin{center}
 \includegraphics[width=7cm]{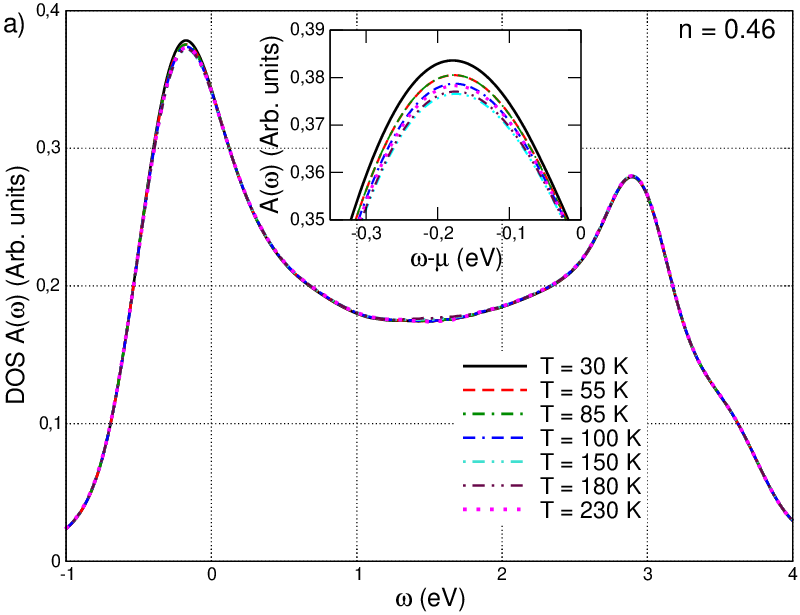}
 \includegraphics[width=5.0cm]{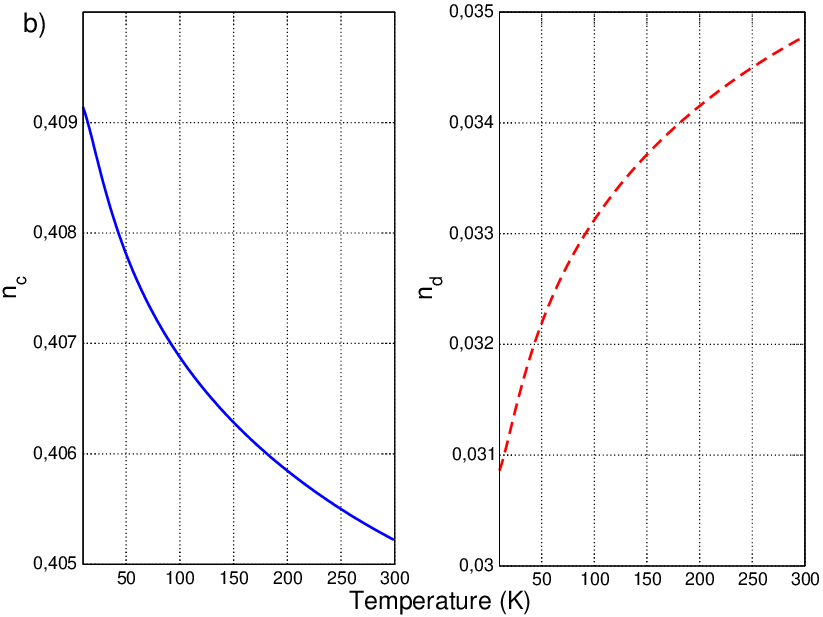}
    \caption[]{ Temperature dependence at $n=0.46$. (a) DOS at different temperatures (as indicated in the figure). Inset: amplification, showing the 
    T-dependence of the peak near $E_F$. 
    (b) $c$ and $d$ band fillings, as a function of temperature.  Parameters: $U=1.5 eV = 2 V $. Other parameters as in Fig.1, except for:  $ \nu = 9 $.}
  \label{ef-T-Aw}
  \end{center}
  \end{figure}

First, we will present our prediction for the temperature dependence of the total density of states of LaO$_{1-x}$F$_{x}$BiS$_{2}$ compounds.
  In Figure \ref{ef-T-Aw}(a), we show our calculated DOS  for several temperatures at $n = 0.46$ ( doping value of sample studied by ARPES 
  at $T = 10 K $\cite{terashima2014}). 
 
  Notice that the total DOS in Figure \ref{ef-T-Aw}(a), which includes the sum over contributions from the whole BZ, is nearly independent of temperature. The peak near $E_F$ 
  is shown amplified in the figure inset:  increasing temperature a very slight reduction of spectral weight of the peak is observable.   
  In Figure \ref{ef-T-Aw}(b) we further plot the corresponding band fillings, $n_{c}$ and $n_{d}$, as function of temperature. 
    Since for the doping value and temperatures considered, the Fermi level falls inside $c$ and $d$ bands, whose fillings change smoothly, 
    the negligible temperature dependence  of the total DOS is not surprising.  

\begin{figure}[h!]
  \begin{center}
  \includegraphics[width=8.6cm]{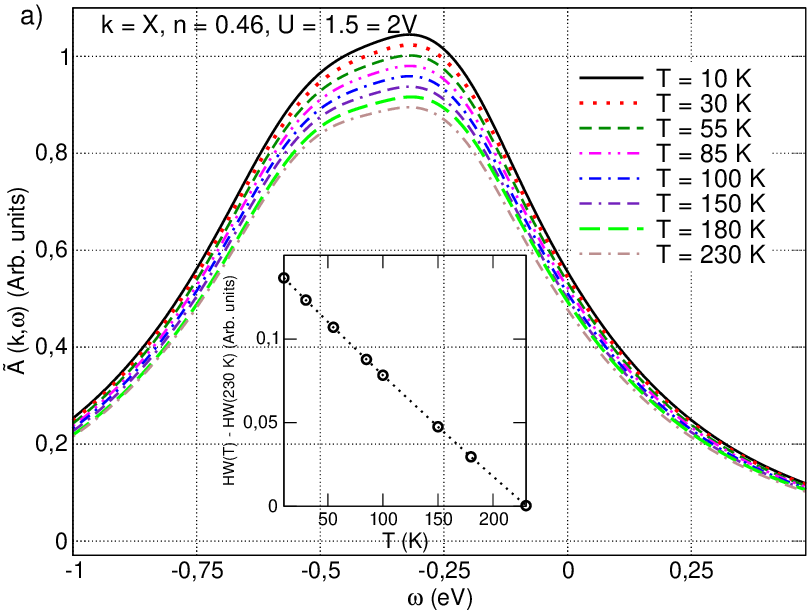}
  \includegraphics[width=8.6cm]{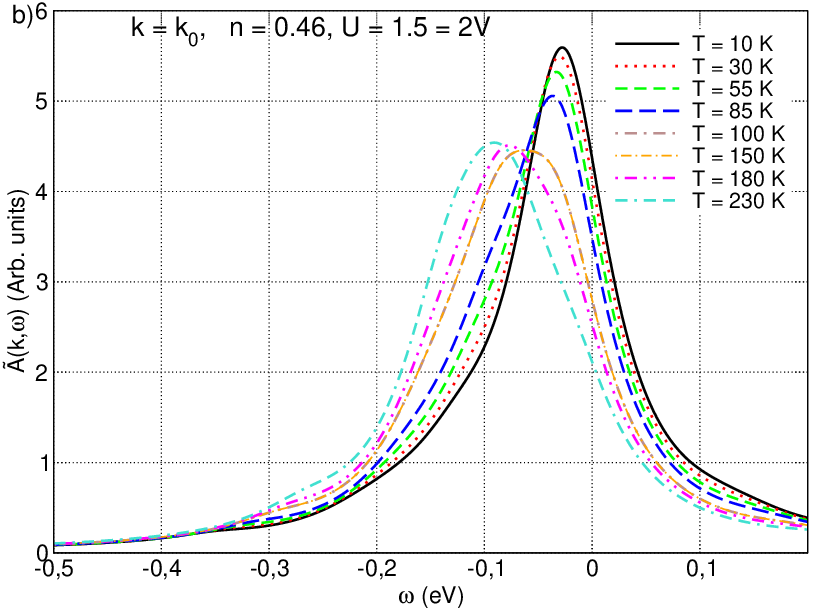}
    \caption[]{ a) Temperature dependence of $\tilde{A}(\vec{k},\omega)$, at $\vec{k}=X$ (for the temperatures indicated in the figure) and b) Temperature dependence of $\tilde{A}(\vec{k},\omega)$, at $\vec{k}_{0}=(0.45\pi,0.45\pi)$. $U = 1.5$ eV, $V=0.75$, $n=0.46$. Other parameters as in Fig.1, except for:  $ \nu = 9 $.}
  \label{ef-T-Akw-kX-La}
  \end{center}
  \end{figure}

Next, we will focus on the temperature dependence of the spectral density function, predicting it for LaO$_{1-x}$F$_{x}$BiS$_{2}$ compounds where 
no T-dependent ARPES results are available. We do make a  comparison with the available temperature-dependent ARPES data for  NdO$_{0.7}$F$_{0.3}$BiS$_{2}$ \cite{zeng2014}.  To allow for direct comparison with ARPES data,\cite{zeng2014} 
the spectral density results presented  in Figure \ref{ef-T-Akw-kX-La},  which are denoted by $ \tilde{A}( \vec{k}, \omega)$, 
were obtained by the convolution of  $ A(  \vec{k}, \omega)$  with the experimental energy resolution $ \delta E = 25$ meV.\cite{zeng2014}

First, we predict the evolution with temperature of the spectral density function for LaO$_{1-x}$F$_{x}$BiS$_{2}$ compounds, 
fixing $\vec{k}$ at the relevant high-symmetry points of the BZ, namely the conduction band minima $X$ and $\vec{k}_{0}$ shown in Fig.\ref{renorbands}.
In Figure \ref{ef-T-Akw-kX-La}(a), we show our calculated temperature dependence for $\tilde{A}(\vec{k} = X ,\omega)$ at $n=0.46$, doping which corresponds to 
LaO$_{0.54}$F$_{0.46}$BiS$_{2}$ compound studied by ARPES only at the temperature $ T= 10 K$ in Ref.\cite{terashima2014}. 
In Figure \ref{ef-T-Akw-kX-La}(a), the spectral density at X  exhibits a broad slightly asymmetric hump, extending from about -1 eV to slightly above $E_F$, 
in agreement with Ref.\cite{terashima2014} at $T = 10 K$ (and our results of Fig.4.b convoluted with $ \delta E = 25$ meV). 
Increasing temperature, we find that the spectral weight is monotonically reduced, and no new peaks appear with temperature. 
The inset in Figure \ref{ef-T-Akw-kX-La}(a), shows the integrated spectral weight of the hump as a function of temperature, with the 
substraction of the integrated spectral weight of the hump at $T=230$ (denoted as $ HW (T) - HW ( 230 K)$ ) as in the data analysis reported for the T-dependent ARPES   
experiment in NdO$_{1-x}$F$_{x}$BiS$_{2}$.\cite{zeng2014} 
In Figure \ref{ef-T-Akw-kX-La}(b), we predict the evolution of $\tilde{A}(\vec{k},\omega)$  with temperature at $\vec{k}_{0}$, not yet probed by ARPES.  
The spectral function exhibits a single peak near $E_F$, which is shifted to lower energies as temperature is increased.  Also, the standard thermal broadening effect with 
redistribution of the peak's spectral weight is visible. We checked that the total integrated spectral weight as a function of temperature remains constant. 
Between 100 and 150 K, we find negligible temperature dependence of $\tilde{A}(\vec{k},\omega)$ at $\vec{k}_{0}$.

The temperature dependence of $\tilde{A}(\vec{k},\omega)$ we are predicting for LaO$_{0.54}$F$_{0.46}$BiS$_{2}$ 
in Figure \ref{ef-T-Akw-kX-La} could be probed by future ARPES experiments  at X and at $\vec{k}_0$. 

\begin{figure}[h!]
  \begin{center}
  \includegraphics[width=8.6cm]{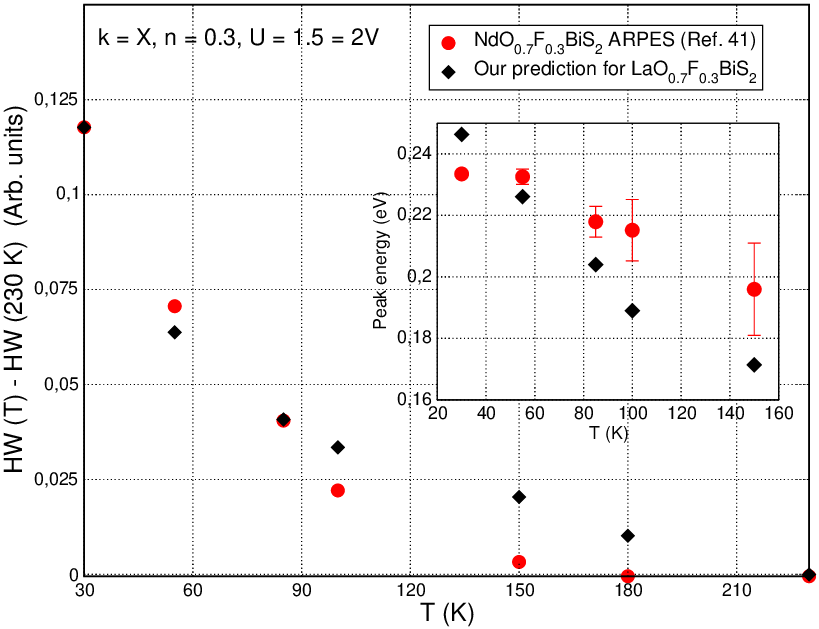}
    \caption[]{ Integrated spectral weight of $A(\vec{k},\omega)$, at $\vec{k}=X$ (temperatures are indicated in the figure).
 Inset: temperature dependence of the energy of the maximum of the hump.    Experimental energy resolution of $25$ meV, $U = 1.5$ eV, $V=0.75$, $n=0.30$. 
Other parameters as in Fig.1, except for:  $ \nu = 9 $.}
  \label{ef-T-Akw-X-Nd}
  \end{center}
  \end{figure}

Finally, in Figure \ref{ef-T-Akw-X-Nd}, we compare the temperature dependent spectral density results we predict for LaO$_{0.7}$F$_{0.3}$BiS$_{2}$, 
with the only available temperature-dependent ARPES experiment we found for a BiS$_2$-based compound:  NdO$_{0.7}$F$_{0.3}$BiS$_{2}$ \cite{zeng2014}. 
This doping level corresponds to the optimal electron doping for superconductivity ( T$_C$ = 4 K) in the NdO$_{1-x}$F$_{x}$BiS$_{2}$ family.
 ( To allow a direct comparison of the results included in Figure \ref{ef-T-Akw-X-Nd}, we have used the same 
 data treatment adopted in  Ref.\cite{zeng2014}: to remove thermal broadening effects, 
 the  spectral data $\tilde{A} (\vec{k},\omega)$ were divided by resolution-convoluted Fermi functions at each temperature,
  and finally all spectra were multiplied by the Fermi function at $T = 10$ K.) 
  
 Concretely, in Figure \ref{ef-T-Akw-X-Nd}, we compare the temperature dependence of the integrated hump weight $ HW (T) - HW ( 230 K)$ for both compounds, 
 and in the inset compare the temperature dependence of the hump positions. 
 Though a monotonous decrease of both magnitudes with temperature 
  is observable, there are differences between our predictions  for LaO$_{0.7}$F$_{0.3}$BiS$_{2}$, and the ARPES results for NdO$_{0.7}$F$_{0.3}$BiS$_{2}$ \cite{zeng2014}.
But these are to be expected, since our approach adds correlations to the effective two-orbital model with tight-binding parameters specifically determined 
for LaO$_{1-x}$F$_{x}$BiS$_{2}$ compounds\cite{usui2012}. 
No similar tight-binding parameters were determined specifically for NdO$_{1-x}$F$_{x}$BiS$_{2}$ compounds yet, and differences between the band structures 
are expected, according to DFT calculations\cite{wang2014}: while for LaO$_{1-x}$F$_{x}$BiS$_{2}$ the two band minima closest to $E_F$ 
(the $d$-band minimum at $X$ and the $c$-band  minimum at $\vec{k}_{0}$) have very similar energies, 
this is not the case for NdO$_{1-x}$F$_{x}$BiS$_{2}$ where for $x=0$ an energy difference of 0.8 eV was determined between the two relevant band minima by DFT\cite{wang2014}.

\section{Conclusions}\label{Conclusions}

We have studied the normal state spectral properties of LaO$_{1-x}$F$_{x}$BiS$_{2}$ superconductors,  using an extended Hubbard model based on two correlated effective orbitals and an  analytical approximation to decouple and solve the equations of motion
 for the electron Green's functions. Our results demonstrate that the inclusion of moderate electron correlations in LaO$_{1-x}$F$_{x}$BiS$_{2}$ improves the  description of the available experimental results for these compounds. Concretely, including electron correlations with our second-order perturbative approach 
 we could: i) improve the description of the chemical potential shift of  approximately  0.3 eV between the $x=0$ and $x=0.5$ compounds reported 
 in soft X-ray photoemission experiments at  $ T = 300 K$, by including moderate intra- and inter-orbital correlations: $U=1.5$ eV, and $V=0.75$ eV;  ii) describe the momentum dependence measured in ARPES around  $X$;  iii) and study the  Fermi-surface topological change, which is independent of temperature: it occurs at $x=0.44$ in the absence of correlations, while moderate correlations shift it to lower doping values.
 
 Furthermore,  our analytical approximation has the advantage of enabling us to describe temperature and k-dependent results,  
 which allowed us to explore the spectral density at other BZ regions and in particular predict the momentum dependence around the second relevant 
 band structure minimum located at $\vec{k}_{0}=(0.45\pi,0.45\pi)$.  We also found  that electron correlations enhance the nesting properties 
 of the Fermi surface. Regarding the temperature dependence of the spectral function, our work predicts results which depend on the BZ point: 
 at $X$ we found that the position of the main hump near the Fermi level is unchanged, while an important hump spectral weight reduction  takes place when temperature is increased. Instead  at $k_{0}$  we found that temperature induces a shift of the position of the peak close to the Fermi level, while a spectral weight redistribution which leaves the total hump weight almost constant takes place.  It would be interesting to have our predictions tested, by temperature dependent ARPES experiments in LaO$_{1-x}$F$_{x}$BiS$_{2}$.
  
Our calculated Green's functions could be used to evaluate  other low energy normal state physical properties of LaO$_{1-x}$F$_{x}$BiS$_{2}$: 
being our analytical approach especially useful for analyzing  k-dependent results,  in contrast to other theoretical techniques where self-energies are local.

\section{Acknowledgments}

C.I.V. is Investigadora Cient{\'{\i}}fica of CONICET (Argentina). J.D.Q.F. has a fellowship from CONICET. C.I.V. acknowledges support from CONICET (PIP 0702) and ANPCyT (PICT Raices'2012, nro.1069). J.D.Q.F. acknowledges the hospitality of IIP (International Institute of Physics, Natal), and acknowledges stimulating discussions with G. Martins.

\appendix 

\vspace{0.8cm}

\section[Analytical approximation]{Analytical results for the $c$ and $d$ electron Green's functions.}\label{apen1}

\subsection{First order solution of EOM: Hartree-Fock}

 Below, we state the Hartree-Fock solutions obtained for the  $c$ and $d$ electron Green's functions, as mentioned in Section  \ref{calculations}:

 \begin{equation}
G^{H.F.}_{\sigma}(k, \omega)  \cong \frac{1}{  2\pi\left[  \omega -  E_{c}(k)  - \left(U\langle {n}_{c,\overline{\sigma}}\rangle -  V\langle {n}_{d}\rangle \right)    \right]  } 
\end{equation}

\noindent and,

\begin{equation}
F^{H.F.}_{\sigma}(k, \omega)  \cong \frac{1}{  2\pi\left[  \omega -  E_{d}(k)  - \left(U\langle {n}_{d,\overline{\sigma}}\rangle -  V\langle {n}_{c}\rangle \right)    \right]  } 
\end{equation}

\subsection{Second-order solution of EOM}

Below, we present the  solution in second-order of perturbations on the electron correlations  $U$ and $V$  for 
 $G(\vec{k},\omega)$ and $F(\vec{k},\omega)$ obtained through the analytical approximation\cite{condmat2015}  
 described in Section \ref{calculations}:

 \begin{widetext}
\begin{equation}\label{G}
G_{{\sigma}}(k, \omega) \cong \frac{ \frac{1}{2\pi} + \sum_{k_{2}}  {\left \{ \frac{U}{N}\left(\frac{Y_{1}}{X_{1}}\right)  + \frac{V}{N}\left[ \frac{Y_{2}}{X_{2}} + [A_{3}]_{k_{1}} +  [C_{3}]_{k_{1}}\left(\frac{Y_{1}}{X_{1}}+\frac{Y_{2}}{X_{2}}\right)     \right]  \right\} }     }{ \omega - E_{c}(k) -\sum_{k_{2}}{\left \{\frac{U}{N}\left(\frac{Z_{1}}{X_{1}}\right)+\frac{V}{N} \left[ \frac{Z_{2}}{X_{2}}+[B_{3}]_{k_{1}} +[C_{3}]_{k_{1}} \left( \frac{Z_{1}}{X_{1}}+\frac{Z_{2}}{X_{2}} \right) \right] \right\}} }
\end{equation}

\noindent To abbreviate, we use the following notation for summations: $[ \alpha ]_{k_{1}} \equiv \sum_{k_{1}}{ \alpha }$, for different $\alpha$ coefficients. Meanwhile: 

 \end{widetext}
 
\begin{eqnarray*}
X_{1}(k_{2},k)  = &\left(1-\frac{U}{N}[C_{2}]_{k_{1}}[C_{3}]_{k_{1}}\right) \left( 1-[C_{1}]_{k_{1}}[C_{3}]_{k_{1}}  \right) \\
  &- [C_{1}]_{k_{1}}\left( \frac{U}{N}[C_{3}]_{k_{1}} + 1 \right) \left( \frac{U}{N}[C_{3}]_{k_{1}} + \frac{V}{N} \right) [C_{2}]_{k_{1}}\\
Y_{1}(k_{2},k)  = &\left(1-\frac{U}{N}[C_{2}]_{k_{1}}[C_{3}]_{k_{1}}\right) \left( [A_{1}]_{k_{1}} +  [A_{3}]_{k_{1}}[C_{1}]_{k_{1}} \right)\\
 & + \left( [A_{2}]_{k_{1}} + \frac{U}{N}[C_{2}]_{k_{1}}[A_{3}]_{k_{1}} \right) [C_{1}]_{k_{1}} \left( \frac{U}{N}[C_{3}]_{k_{1}} + 1 \right)\\
Z_{1}(k_{2},k)  = &\left(1-\frac{U}{N}[C_{2}]_{k_{1}}[C_{3}]_{k_{1}}\right)\left( [B_{1}]_{k_{1}} +  [B_{3}]_{k_{1}}[C_{1}]_{k_{1}} \right) \\
& + [B_{2}]_{k_{1}} + \frac{U}{N}[C_{2}]_{k_{1}}[B_{3}]_{k_{1}}
\end{eqnarray*}

\begin{eqnarray*}
 X_{2}(k_{2},k) = &\left(1-[C_{1}]_{k_{1}}[C_{3}]_{k_{1}}\right) \left( 1-\frac{U}{N}[C_{2}]_{k_{1}}[C_{3}]_{k_{1}}  \right)\\
  & - \left(\frac{U}{N}[C_{3}]_{k_{1}} + V\right)[C_{2}]_{k_{1}}  [C_{1}]_{k_{1}} \left([C_{3}]_{k_{1}} + 1\right)\\
 Y_{2}(k_{2},k) = &\left(1-[C_{1}]_{k_{1}}[C_{3}]_{k_{1}}\right) \left( [A_{2}]_{k_{1}} + \frac{U}{N} [C_{2}]_{k_{1}}[A_{3}]_{k_{1}} \right)\\
   &  + \left( \frac{U}{N} [C_{3}]_{k_{1}} + V \right) [C_{2}]_{k_{1}}\left( [A_{1}]_{k_{1}} +[A_{3}]_{k_{1}}[C_{1}]_{k_{1}}\right)\\
 Z_{2}(k_{2},k) =& \left(1-[C_{1}]_{k_{1}}[C_{3}]_{k_{1}}\right) \left( [B_{2}]_{k_{1}} + \frac{U}{N} [C_{2}]_{k_{1}}[B_{3}]_{k_{1}} \right) \\
  & + \left( \frac{U}{N} [C_{3}]_{k_{1}} + V \right) [C_{2}]_{k_{1}}\left( [B_{1}]_{k_{1}} +[B_{3}]_{k_{1}}[C_{1}]_{k_{1}}\right)
\end{eqnarray*}


\begin{align}
& A_{1}   =   \frac{\langle {n}_{{k}_{1},\overline{\sigma}} \rangle}{2\pi \left[ \omega - E_{c}(k_{2})-U(n_{c}+1) \right]} \nonumber \\
& A_{2}   =   \frac{\langle {N}_{{k}_{1},{\sigma}} \rangle}{2\pi \left[ \omega - E_{c}(k_{2})-Un_{c}-2n_{d}(U+V) + V \right]} \nonumber\\
&A_{3}  =   \frac{\langle {N}_{{k}_{1},\overline{\sigma}} \rangle}{2\pi \left[ \omega - E_{c}(k_{2})-Un_{c}-2Vn_{d} - V \right]} \nonumber
\end{align}

\begin{align}
&B_{1}   =    \frac{\{ 2V\langle {n}_{{k}_{1},\overline{\sigma}}\rangle n_{d} + (2+\langle {n}_{{k}_{1},\overline{\sigma}} \rangle)U{n}_{c}+ U(\langle {n}_{{k}_{1},\overline{\sigma}} \rangle - \langle {n}_{{k}_{1}-{k}_{2}+k,\overline{\sigma}} \rangle)\}}{\omega - \omega_{1}} \nonumber\\
&B_{2}   =   \frac{\{ (U-V)\langle {N}_{{k}_{1},{\sigma}}\rangle + 2V\langle {N}_{{k}_{1},{\sigma}} \rangle n_{d}- V(\langle {N}_{{k}_{1},{\sigma}} \rangle - \langle N_{{k}_{1}-{k}_{2}+k,{\sigma}} \rangle) +V \}}{\omega - \omega_{2}} \nonumber \\
&B_{3}   =   \frac{\{ U\langle {N}_{{k}_{1},\overline{\sigma}}\rangle {n}_{c} + V(1 - \langle {n}_{{k}_{1},\overline{\sigma}} \rangle -  \langle {n}_{{k}_{1}-{k}_{2}+k,\overline{\sigma}} \rangle)n_{d} + V(\langle {n}_{{k}_{1},\overline{\sigma}} \rangle \}}{\omega - \omega_{3}}\nonumber
\end{align}

\begin{align}
&C_{1}   =    \frac{\frac{V}{N} \left[\langle {n}_{{k}_{1},\overline{\sigma}} \rangle - \langle {n}_{{k}_{1}-{k}_{2}+k,\overline{\sigma}} \rangle \right]}{\omega - \omega_{1}} \nonumber\\
&C_{2}   =   \frac{\left[\langle {N}_{{k}_{1},{\sigma}} \rangle - \langle {N}_{{k}_{1}-{k}_{2}+k,{\sigma}} \rangle \right]}{\omega - \omega_{2}} \nonumber\\
&C_{3}   =   \frac{\left[\langle {N}_{{k}_{1},\overline{\sigma}} \rangle - \langle {N}_{{k}_{1}-{k}_{2}+k,\overline{\sigma}} \rangle \right] }{\omega - \omega_{3}} \nonumber
\end{align}


\begin{align}
\,\,\,\, \,\,\,\, &\omega_{1}   =   \left[ {E}_{c}({k}_{1}-{k}_{2}+k) - {E}_{c}({k}_{1}) + {E}_{c}({k}_{2}) + U({n}_{c} + 1) \right]\nonumber \\
&\omega_{2}   =   \left[ {E}_{d}({k}_{1}-{k}_{2}+k) - {E}_{d}({k}_{1}) + {E}_{c}({k}_{2}) + U{n}_{c}  \right. \nonumber\\
& \left. + 2n_{d}(U+V) - V \right] \nonumber\\
&\omega_{3}   =    \left[ {E}_{d}({k}_{1}-{k}_{2}+k) - {E}_{d}({k}_{1}) + {E}_{c}({k}_{2}) + U{n}_{c} + 2Vn_{d} + V \right]\nonumber
\end{align}

 \begin{widetext}
Analogously,   for  $F_{\sigma}(k,\omega)$ corresponding to the d-electrons we obtained:

\begin{equation}\label{F}
F_{{\sigma}}(k, \omega) \cong \frac{ \frac{1}{2\pi} + \sum_{k_{2}}  {\left \{ \frac{U}{N}\left(\frac{Y^{*}_{1}}{X^{*}_{1}}\right)  + \frac{V}{N}\left[ \frac{Y^{*}_{2}}{X^{*}_{2}} + [A^{*}_{3}]_{k_{1}}  +  [C^{*}_{3}]_{k_{1}}\left(\frac{Y^{*}_{1}}{X^{*}_{1}}+ \frac{Y^{*}_{2}}{X^{*}_{2}}\right)     \right]  \right\} }     }{ \omega - E_{d}(k) -\sum_{k_{2}}{\left \{\frac{U}{N}\left(\frac{Z^{*}_{1}}{X^{*}_{1}}\right)+\frac{V}{N} \left[ \frac{Z^{*}_{2}}{X^{*}_{2}}+[B^{*}_{3}]_{k_{1}} +[C^{*}_{3}]_{k_{1}} \left( \frac{Z^{*}_{1}}{X^{*}_{1}}+\frac{Z^{*}_{2}}{X^{*}_{2}} \right) \right] \right\}} }
\end{equation}
\end{widetext}

\noindent where:

\begin{eqnarray*}
X^{*}_{1}(k_{2},k) = &\left(1-\frac{U}{N}[C^{*}_{2}]_{k_{1}}[C^{*}_{3}]_{k_{1}}\right) \left( 1-[C^{*}_{1}]_{k_{1}}[C^{*}_{3}]_{k_{1}}  \right) \\
& - [C^{*}_{1}]_{k_{1}}\left( \frac{U}{N}[C^{*}_{3}]_{k_{1}} + 1 \right) \left( \frac{U}{N}[C^{*}_{3}]_{k_{1}} + \frac{V}{N} \right) [C^{                                                                        *}_{2}]_{k_{1}}\\
Y^{*}_{1}(k_{2},k)  = &\left(1-\frac{U}{N}[C^{*}_{2}]_{k_{1}}[C^{*}_{3}]_{k_{1}}\right) \left( [A^{*}_{1}]_{k_{1}} +  [A^{*}_{3}]_{k_{1}}[C^{*}_{1}]_{k_{1}} \right) \\
& + \left( [A^{*}_{2}]_{k_{1}} + \frac{U}{N}[C^{*}_{2}]_{k_{1}}[A^{*}_{3}]_{k_{1}} \right) [C^{*}_{1}]_{k_{1}}\left( \frac{U}{N}[C^{*}_{3}]_{k_{1}} + 1 \right)\\
Z^{*}_{1}(k_{2},k)  = &\left(1-\frac{U}{N}[C^{*}_{2}]_{k_{1}}[C^{*}_{3}]_{k_{1}}\right)\left( [B^{*}_{1}]_{k_{1}} +  [B^{*}_{3}]_{k_{1}}[C^{*}_{1}]_{k_{1}} \right) \\
& + [B^{*}_{2}]_{k_{1}} + \frac{U}{N}[C^{*}_{2}]_{k_{1}}[B^{*}_{3}]_{k_{1}}
\end{eqnarray*}

\begin{align}
&A^{*}_{1}  =   \frac{\langle {N}_{{k}_{1},\overline{\sigma}} \rangle}{2\pi \left[ \omega - E_{d}(k_{2})-U(n_{d}+1) \right]} \nonumber\\
&A^{*}_{2}   =   \frac{\langle {n}_{{k}_{1},{\sigma}} \rangle}{2\pi \left[ \omega - E_{d}(k_{2})-Un_{d}-2n_{c}(U+V) + V \right]} \nonumber\\
&A^{*}_{3}   =   \frac{\langle {n}_{{k}_{1},\overline{\sigma}} \rangle}{2\pi \left[ \omega - E_{d}(k_{2})-Un_{d}-2Vn_{c} - V \right]}\nonumber
\end{align}

\begin{align}
&B^{*}_{1}   =    \frac{\{ 2V\langle {N}_{{k}_{1},\overline{\sigma}}\rangle{n}_{c} + (2+\langle {N}_{{k}_{1},\overline{\sigma}} \rangle)Un_{d}+ U(\langle {N}_{{k}_{1},\overline{\sigma}} \rangle - \langle {N}_{{k}_{1}-{k}_{2}+k,\overline{\sigma}} \rangle)\}}{\omega - \omega^{*}_{1}} \nonumber\\
&B^{*}_{2}   =   \frac{\{ (U-V)\langle {N}_{{k}_{1},{\sigma}}\rangle + 2V\langle {n}_{{k}_{1},{\sigma}} \rangle {n}_{c}- V(\langle {n}_{{k}_{1},{\sigma}} \rangle - \langle n_{{k}_{1}-{k}_{2}+k,{\sigma}} \rangle) +V \}}{\omega - \omega^{*}_{2}} \nonumber\\
&B^{*}_{3}   =   \frac{\{ U\langle {n}_{{k}_{1},\overline{\sigma}}\rangle n_{d} + V(1 - \langle {N}_{{k}_{1},\overline{\sigma}} \rangle -  \langle {N}_{{k}_{1}-{k}_{2}+k,\overline{\sigma}} \rangle){n}_{c} + V(\langle {N}_{{k}_{1},\overline{\sigma}} \rangle \}}{\omega - \omega^{*}_{3}}\nonumber
\end{align}

\begin{eqnarray*}
 X^{*}_{2}(k_{2},k) = &\left(1-[C^{*}_{1}]_{k_{1}}[C^{*}_{3}]_{k_{1}}\right) \left( 1-\frac{U}{N}[C^{*}_{2}]_{k_{1}}[C^{*}_{3}]_{k_{1}}  \right)\\
 & - \left(\frac{U}{N}[C^{*}_{3}]_{k_{1}} + V\right)[C^{*}_{2}]_{k_{1}}  [C^{*}_{1}]_{k_{1}} \left([C^{*}_{3}]_{k_{1}} + 1\right)\\
 Y^{*}_{2}(k_{2},k) = &\left(1-[C^{*}_{1}]_{k_{1}}[C^{*}_{3}]_{k_{1}}\right) \left( [A^{*}_{2}]_{k_{1}} + \frac{U}{N} [C^{*}_{2}]_{k_{1}}[A^{*}_{3}]_{k_{1}} \right) \\
 & + \left( \frac{U}{N} [C^{*}_{3}]_{k_{1}} + V \right) [C^{*}_{2}]_{k_{1}}\left( [A^{*}_{1}]_{k_{1}} +[A^{*}_{3}]_{k_{1}}[C^{*}_{1}]_{k_{1}}\right)\\
 Z^{*}_{2}(k_{2},k) =& \left(1-[C^{*}_{1}]_{k_{1}}[C^{*}_{3}]_{k_{1}}\right) \left( [B^{*}_{2}]_{k_{1}} + \frac{U}{N} [C^{*}_{2}]_{k_{1}}[B_{3}]_{k_{1}} \right)\\
 &  + \left( \frac{U}{N} [C^{*}_{3}]_{k_{1}} + V \right) [C^{*}_{2}]_{k_{1}}\left( [B^{*}_{1}]_{k_{1}} +[B^{*}_{3}]_{k_{1}}[C^{*}_{1}]_{k_{1}}\right)
\end{eqnarray*}

\begin{align}
&C^{*}_{1}  =   \frac{V}{N} \frac{ \left[\langle {N}_{{k}_{1},\overline{\sigma}} \rangle - \langle {N}_{{k}_{1}-{k}_{2}+k,\overline{\sigma}} \rangle \right]}{\omega - \omega^{*}_{1}} \nonumber\\
&C^{*}_{2}   =   \frac{\left[\langle {n}_{{k}_{1},{\sigma}} \rangle - \langle {n}_{{k}_{1}-{k}_{2}+k,{\sigma}} \rangle \right]}{\omega - \omega^{*}_{2}} \nonumber\\
&C^{*}_{3}   =   \frac{\left[\langle {n}_{{k}_{1},\overline{\sigma}} \rangle - \langle {n}_{{k}_{1}-{k}_{2}+k,\overline{\sigma}} \rangle \right] }{\omega - \omega^{*}_{3}}\nonumber
\end{align}


\begin{align}
\,\,\,\, \,\,\,\, &\omega^{*}_{1}   =    \left[ {E}_{d}({k}_{1}-{k}_{2}+k) - {E}_{d}({k}_{1}) + {E}_{d}({k}_{2}) + U(n_{d} + 1) \right] \nonumber\\
&\omega^{*}_{2}   =   \left[ {E}_{c}({k}_{1}-{k}_{2}+k) - {E}_{c}({k}_{1}) + {E}_{d}({k}_{2}) + Un_{d}  \right. \nonumber\\
& \left. + 2{n}_{c}(U+V) - V \right] \nonumber\\
&\omega^{*}_{3}   =    \left[ {E}_{c}({k}_{1}-{k}_{2}+k) - {E}_{c}({k}_{1}) + {E}_{d}({k}_{2}) + Un_{d} + 2V{n}_{c} + V \right]\nonumber
\end{align}



\begin{thebibliography}{99}

\bibitem{mizuguchi2012} Y. Mizuguchi,  et al.,  Phys. Rev. B 86, 220510(R) (2012).
\bibitem{awana-SSC2013} V.P.S. Awana, Anuj Kumar, Rajveer Jha,  et al., Solid State Commun. 157, 21 (2013).
\bibitem{review2015} D. Yazici, et al., Physica C (2015), doi: http://dx.doi.org/10.1016/j.physc.2015.02.025, and references therein.
\bibitem{review2014}Y. Mizuguchi, J. Phys. Chem. Solids (2014), http://dx.doi.org/10.1016/j.jpcs.2014.09.003i, and references therein.
 
\bibitem{awana-JPCS2015} Rajveer Jha, H. Kishan and V.P.S. Awana, J. Phys. Chem. Solids 84, 17 (2015).


\bibitem{maziopa2014} A. Krzton-Maziopa, et al., J. Phys.: Condens. Matter 26, 215702 (2014).
\bibitem{li2013} B. Li, et al.,  Europhys. Lett. 101, 47002 (2013).
\bibitem{deguchi2013} K. Deguchi, et al., Europhys. Lett. 101, 17004 (2013).
\bibitem{cruz2008} C. Cruz et al, Nature 453, 899 (2008).
\bibitem{xing2012} J. Xing, et al., Phys. Rev. B 86 214518 (2012).
 \bibitem{demura2015} S. Demura, et al, J. Phys. Soc. Jpn. 84, 024709 (2015).
 \bibitem{nagao2015} M. Nagao,  cond-mat preprint: 1511.00219v1 (2015). (To be published in: Novel Superconducting Materials)
\bibitem{usui2012} H. Usui, K. Suzuki, K. Kuroki, Phys. Rev. B 86, 220501 (2012).
\bibitem{wan2013} Z. Wan, et al., Phys. Rev. B 87, 115124 (2013).
\bibitem{liu2013} S. L. Liu,  J. Superc. Novel Mag. 26, 3411 (2013).
\bibitem{shein2013} I. R. Shein, A. L. Ivanovskii, JETP Letters 96,769 (2013).
\bibitem{suzuki2013} K. Suzuki, H. Usui, K. Kuroki, Phys. Procedia 45, 21 (2013), proceedings of the 25th International Symposium on
Superconductivity (ISS2012).
\bibitem{gao2014} Yi Gao, et al., Phys. Rev. B 90, 054518 (2014).
\bibitem{morice2015}  C. Morice,  et al., J. Phys.: Condens. Matter 27, 135501 (2015).
\bibitem{wu2014} X. Wu, et al.,   arXiv:1403.5949 (2014).
\bibitem{yang2013} Y. Yang, et al., Phys. Rev. B 88, 094519 (2013).
\bibitem{zhou2013} T. Zhou and Z. D. Wang, J. Supercond. Nov. Magn. 26, 2735 (2013).
\bibitem{terashima2014} K. Terashima, et al, Phys. Rev. B, 90, 220512(R) (2014).
\bibitem{nagira2014} S. Nagira, et al.,  J. Phys. Soc. Jpn. 83, 033703 (2014).
\bibitem{yildirim2013} T. Yildirim, Phys. Rev. B 87, 020506 (2013).
\bibitem{martins2013} G. B. Martins, A. Moreo, E. Dagotto, Phys. Rev. B 87, 081102 (2013).
\bibitem{liang2013} Y. Liang, et al., Frontiers Phys. 9, 47002 (2013).
\bibitem{yao2015} H. Yao, F. Yang, Phys. Rev. B 92, 035132 (2015).
\bibitem{liang2014} Yi Liang,  et al., Front. Phys., 9, 194 (2014).
\bibitem{condmat2015} J. D. Querales Flores, C. I. Ventura,  R. Citro  and J.J. Rodr\'iguez-N\'u\~nez, cond-mat preprint: 1502.08042v3 (2015).
\bibitem{raghu2008} S. Raghu, et al., Phys. Rev. B 77, 220503(R) (2008).
\bibitem{zubarev} D. N. Zubarev, Sov. Phys. Usp. 3, 320 (1960). 
\bibitem{nolting} W. Nolting, \textit{ Fundamentals of Many-Body Physics}, translated by W. D. Brewer (Springer - Verlag Berlin Heidelberg 2009).
\bibitem{roth1969} L. M. Roth, Phys. Rev. 184, 451 (1969).
\bibitem{macot} L. Macot and B. Frank, Phys. Rev. B 41, 4469 (1990).
\bibitem{scalapino2012} D. J. Scalapino, Rev. Mod. Phys. 84, 1383 (2012).
\bibitem{dagotto2011} E. Dagotto et al., Front. Phys. 6(4), 379 (2011).
\bibitem{hague} J. P. Hague, J. Phys. Condens. Matter 17, 1387 (2005).
\bibitem{yamada} K. Yamada, Electron Correlation in Metals (Cambridge University Press, Cambridge, 2004).
\bibitem{ye2014} Z. R. Ye, et al., Phys. Rev. B 90  045116 (2014).
\bibitem{zeng2014} L. K. Zeng, et al., Phys. Rev. B 90, 054512 (2014).
\bibitem{wang2014} X. B. Wang, et al., Phys. Rev. B 90,  054507 (2014).







\end{thebibliography}
\end{document}